\documentclass[twocolumn, a4paper, 9pt]{article}

\usepackage{subcaption}
\usepackage{utfsym}
\usepackage{lineno}
\usepackage{fourier}
\usepackage{relsize}
\usepackage{hyperref}
\usepackage[english]{babel}
\usepackage{booktabs}
\usepackage{amsmath}
\usepackage{multirow}
\usepackage[left=40pt,right=40pt,top=52pt,bottom=50pt]{geometry}
\usepackage{authblk}
\usepackage{balance}
\usepackage{multicol}
\usepackage{natbib}
\newcommand{\figref}[1]{Fig. \ref{#1}}
\newcommand{\tabref}[1]{Tab. \ref{#1}}

\begin{document}

\title{Deep learning meets tree phenology modeling: PhenoFormer vs. process-based models}





\author[1]{Vivien Sainte Fare Garnot}
\author[2,3]{Lynsay Spafford}
\author[3]{Jelle Lever}
\author[4]{Christian Sigg}
\author[4]{Barbara Pietragalla}
\author[3]{Yann Vitasse}
\author[3]{Arthur Gessler}
\author[1]{Jan Dirk Wegner}




\affil[1]{Department of Mathematical Modeling and Machine Learning (DM3L), University of Zurich, Switzerland}

\affil[2]{St. Francis Xavier University, Nova Scotia, Canada}

\affil[3]{Swiss Federal Institute for Forest, Snow and Landscape Research (WSL), Birmensdorf, Switzerland}

\affil[4]{Swiss Federal Office of Meteorology and Climatology (MeteoSwiss), Zurich, Switzerland}

\date{}






\twocolumn[
\maketitle
\abstract{
Phenology, the timing of cyclical events of plant life such as leaf emergence and colouration, is a key phenomenon in the bio-climatic system. Driven by climate change, shifts in the timing of phenological events have far reaching consequences on the ecosystems and on the climate system itself. To anticipate the impacts of different climate change scenarios, it is therefore crucial to design accurate phenology models that can predict the date of occurrence of a phenological phase
even when climatic conditions change significantly from historical ones. 
Existing approaches either rely on hypothesis-driven process models or data-driven statistical methods. Process-based models encompass several dormancy stages and integrate different phenology drivers, while statistical approaches use simple linear or traditional machine learning models. Several studies have shown that statistical methods are outperformed by process models when predicting in external conditions, \textit{i.e.}, when predicting on climatic conditions outside of the range of the training distribution, typically when using climate change scenarios. However, in the realm of statistical methods, deep learning approaches have yet to be explored for  climate phenology modeling.
Here, we introduce a neural architecture, dubbed PhenoFormer, and show that it is better suited than traditional statistical methods at predicting phenology under shift in climate data distribution, while also bringing significant improvements or performing on par to the best performing process-based models. 
We conduct our numerical experiments on a country-scale dataset spanning $70$ years of climatic data and around $70,000$ phenological observations of $9$ woody plant species for leaf emergence and colouration in Switzerland. We extensively compare PhenoFormer to $18$ different process-based models and $3$ traditional machine learning baselines in different distribution shift configurations. 
When predicting in external conditions, the average performance across species of PhenoFormer improves over the best traditional machine learning method by $13\%$ R2 and $1.1$ day RMSE for spring phenology, and $11\%$ R2 and $0.7$ day RMSE for autumn phenology, while also performing on par or better than the best process-based models. 
Our results demonstrate that deep learning has the potential to be a valuable methodological tool for accurate climate-phenology prediction, and our PhenoFormer is a first promising step in improving phenological predictions before a complete understanding of the underlying physiological mechanisms is available.  
}
]

\clearpage

\section{Introduction}
Phenology is the study of the cyclic phenomena occurring in plant and animal life. In particular, plant phenology focuses on the timing of different phases of the plant life cycle such as the development of new leaves in spring and their colouration in autumn, and studies the mechanisms and factors driving these phenomena. Climate conditions are one of the main drivers of plant phenology. In fact, over the past decades, one of the most salient impacts of climate change has been a shift in plant phenology, with earlier leaf emergence \citep{piao2019climatefeedback2}. In turn, changes in plant phenology have far-reaching consequences, affecting the synchronization of ecological interactions \citep{renner2018climate}, as well as feedback effects on the climate system itself \citep{richardson2013climatefeedback1, piao2019climatefeedback2, norton2023climatefeedback3}. Indeed, changes in plant phenology directly impact the length of the season and ecosystem productivity, as well as biophysical vegetation properties. Hence, phenology is considered an essential biodiversity variable \citep{schmeller2018ebv} and is linked to essential climate variables such as biomass and albedo  \citep{bojinski2014ECV}. 
Because of its  ramifications, plant phenology is a key phenomenon to model for the  projection of the impacts of climate change on the biosphere and the climate system. 

One important pathway to make such projections consists in climate-phenology modeling, where plant phenological events are predicted based on time series of weather data. Such models can then be applied to climate change simulations to make projections. 
Phenology models addressing this task can broadly be categorised into eco-physiological or statistical approaches.
Eco-physiological process models are mechanistic models which simulate cause-effect relationships between phenophases and drivers of variation established from experiments, with physically interpretable parameters \citep{chuine2017process}. In contrast, statistical models relate drivers of variation to phenophases without physically interpretable parameters and without explicit cause-effect relationships \citep{asse2020process}.

The simulation of the timing of phenological events with process models offers the opportunity to directly incorporate experimental findings into the prediction of phenology, such as when plants respond to warming spring temperatures, or which temperatures elicit the onset of dormancy. Process models used today vary in complexity, encompassing several successive dormancy stages along with multiple interactive drivers of phenology, such as thermal and daylength cumulative influences \citep{caffarra2011modelling, hufkens2018phenor}.   

While more complex models, including more drivers and dormancy stages, offer improved realism based on experimental findings, large-scale process model intercomparison studies have reported that simpler models, such as the growing degree-day model, often perform similarly well or better than more complex models, with improved spatiotemporal transferability for projective applications \citep{basler2016evaluating, melaas2016multiscale, schadel2023using}. Spring phenology processes have been simulated relatively well across a range of scales, growing conditions, and species \citep{basler2016evaluating, hufkens2018phenor}. In contrast, predictive error is often doubled with autumn process modeling due to knowledge gaps regarding the complex environmental cues of autumn phenology and the higher uncertainty of senescence observations  \citep{bigler2021premature, liu2020modeling, meier2023process, spafford2023climate}. 

Regarding statistical approaches, beyond simple linear models, traditional machine learning models such as Random Forests (RF) and Gradient Boosting Machines (GBM) have been explored to predict plant phenology  \citep{rodriguez2015RFrsPheno,czernecki2018MLpoland, dai2019gbmchina,lee2022RFKoreaColouration,gao2023gaminet}. Such models typically operate on monthly averages of the daily climate time series. In some works, the climate data are complemented with satellite-derived vegetation indices  \citep{dai2019gbmchina}, or local land surface properties  \citep{gao2023gaminet}. 
Several studies have found that such machine learning based approaches can outperform process-based models on spring \citep{dai2019gbmchina} or autumn \citep{gao2023gaminet} phenology when predicting in internal conditions, \textit{i.e.}, with testing data within the same distribution as training data. Yet, future phenology projection poses a challenge of out-of-distribution generalisation, since testing data typically differ significantly from training data because of climate change. 
Few studies have compared the aptitude of process models and statistical approaches for future projection, though \citet{asse2020process} found that simple statistical models show poorer transferability to external conditions than process models.
Similarly, \citet{mo2024advancement} showed that process models outperform statistical approaches when operating under future climate change scenarios. Hence, existing literature suggests that statistical approaches are ill-fitted for phenology projection under climate change compared to process-based models.

\begin{figure*}   
    \begin{subfigure}[b]{0.18\textwidth}
         \centering
         \includegraphics[width=\textwidth]{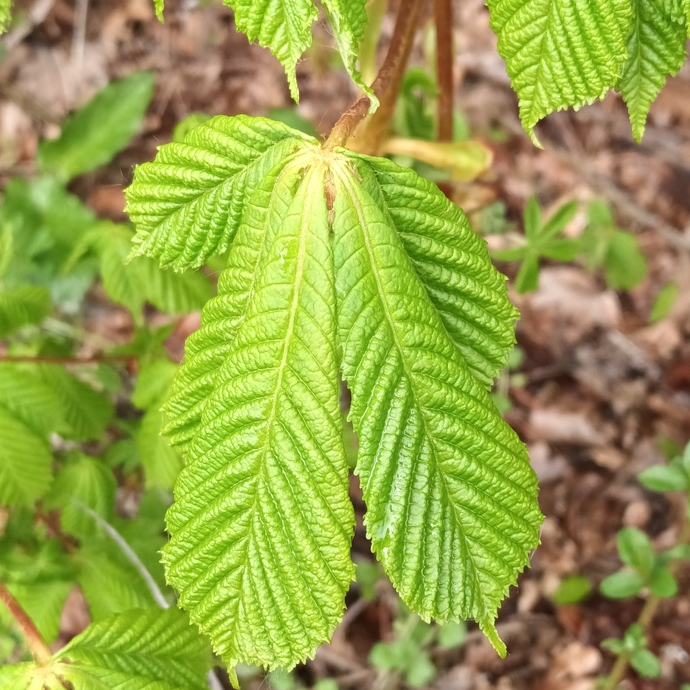}
         \caption{Horse Chestnut}
    \end{subfigure}
    \hfill
    \begin{subfigure}[b]{0.18\textwidth}
         \centering
         \includegraphics[width=\textwidth]{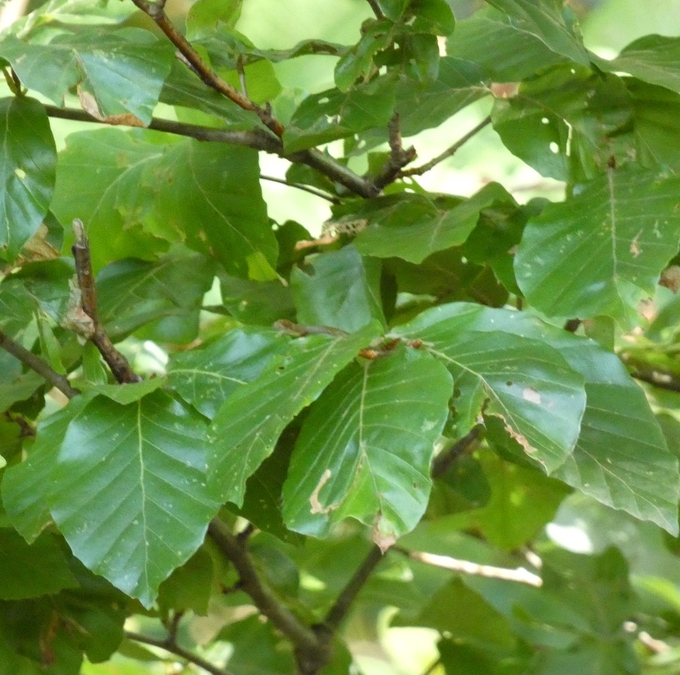}
         \caption{European Beech}
    \end{subfigure}
    \hfill
    \begin{subfigure}[b]{0.18\textwidth}
         \centering
         \includegraphics[width=\textwidth]{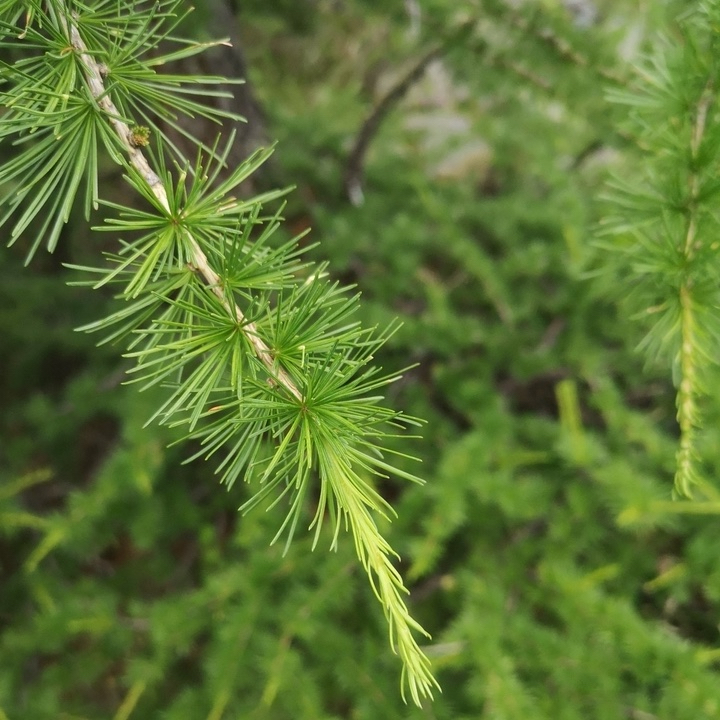}
         \caption{European Larch}
         \label{fig:}
    \end{subfigure}
    \hfill
    \begin{subfigure}[b]{0.18\textwidth}
         \centering
         \includegraphics[width=\textwidth]{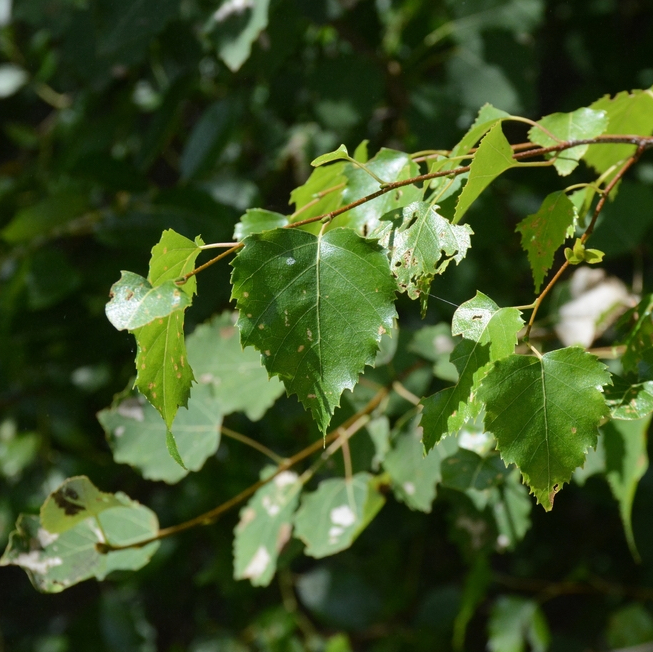}
         \caption{Europ. White Birch}
         \label{fig:}
    \end{subfigure}
    \hfill
    \begin{subfigure}[b]{0.18\textwidth}
         \centering
         \includegraphics[width=\textwidth]{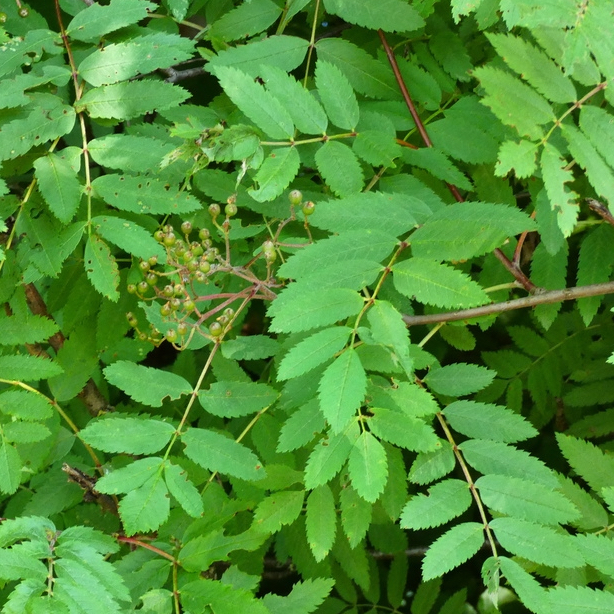}
         \caption{Common Rowan}
         \label{fig:}
    \end{subfigure}
    \vfill
    \begin{subfigure}[b]{0.18\textwidth}
         \centering
         \includegraphics[width=\textwidth]{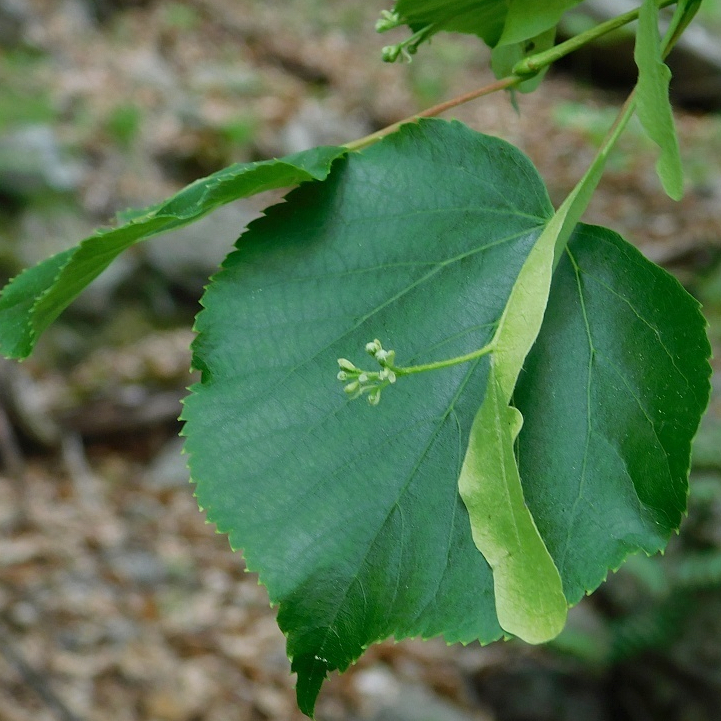}
         \caption{Small Leaved Lime}
         \label{fig:}
    \end{subfigure}
    \hfill
    \begin{subfigure}[b]{0.18\textwidth}
         \centering
         \includegraphics[width=\textwidth]{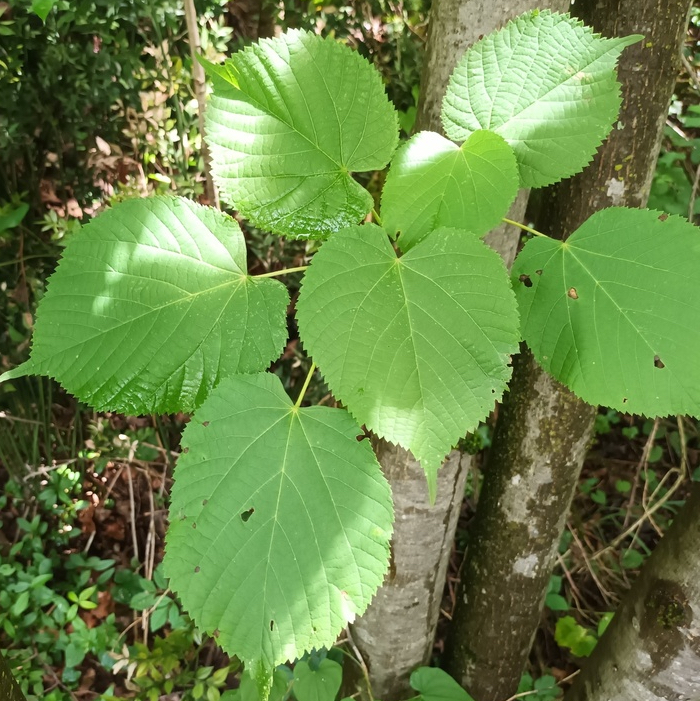}
         \caption{Large Leaved Lime}
         \label{fig:}
    \end{subfigure}
    \hfill
    \begin{subfigure}[b]{0.18\textwidth}
         \centering
         \includegraphics[width=\textwidth]{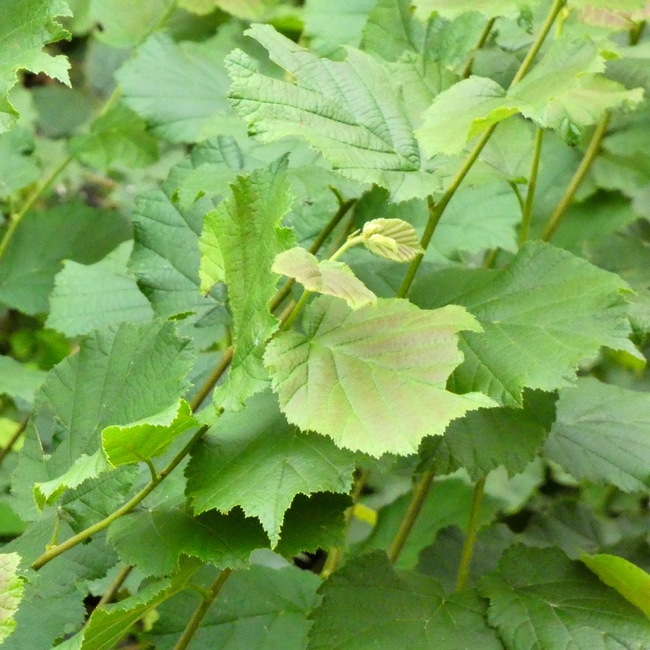}
         \caption{Hazel}
         \label{fig:}
    \end{subfigure}
    \hfill
    \begin{subfigure}[b]{0.18\textwidth}
         \centering
         \includegraphics[width=\textwidth]{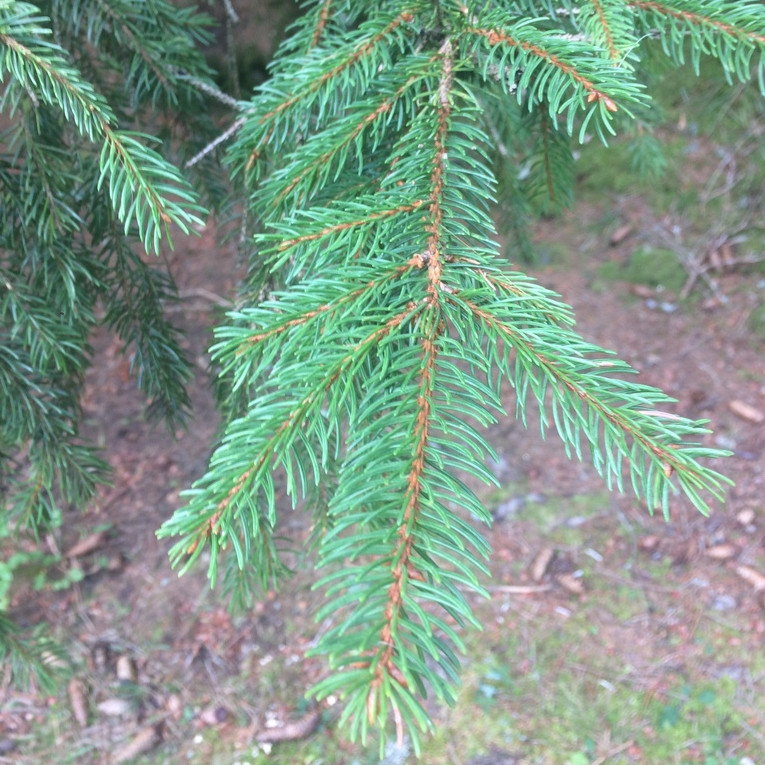}
         \caption{Common Spruce}
         \label{fig:}
    \end{subfigure}
    \caption{Images of the nine species of our dataset. The images are taken during spring months after the unfolding of new leaves or emergence of needles. (Images taken from iNaturalist under CC-0 license.) }
    \label{}
\end{figure*}

However, deep learning, a modern type of statistical learning approach, has yet to be explored for climate phenology modeling.
In the past decade, deep learning research provided a new set of high-performing methodological tools for a variety of data-driven tasks \citep{pichler2023mldleereview}. For phenology in particular, deep learning methods have been used in several studies \citep{katal2022DeepPhenoReview}. Most works use deep learning to process image data, from phenocams \citep{cao2021CNNpheno}, aerial \citep{yang2020uav}, or satellite imagery \citep{xin2020evaluations}. Yet, to the best of our knowledge,  no existing work focuses on predicting phenology from weather records with a deep learning model. Deep learning differs from traditional machine learning approaches in having many more trainable parameters that enable the extraction of learnt features directly from raw inputs instead of pre-computed ones. This superior feature learning capacity of deep learning is typically associated to higher performance on complex tasks and, we believe, can be beneficial to climate-phenology modeling as well. 

In this paper, we propose to explore deep learning based phenology prediction from meteorological time series, and assess in particular its robustness to data distribution shifts compared to process models and traditional machine learning. We design PhenoFormer, a neural architecture based on self-attention \citep{vaswani2017attention}, a type of model particularly well suited to process sequential data. We extensively compare the performance of this deep learning model to three traditional machine learning baselines, as well as 18 different process models for spring and autumn phenology. We focus on predicting leaf unfolding and senescence, as these two phenophases are the basis for carbon sequestration modeling. We conduct our experiments using 9 species of the Swiss Phenology Network \citep{SPN}, a country-scale phenological archive spanning 70 years of history and totalling $67,800$ phenological observations, combined with the DayMetCH meteorological dataset \citep{thornton1997daymetmethods, daymetCH1, daymetCH2}. 
We construct four different dataset splits, corresponding to different distribution shifts between training and testing years, and compare their impacts on the performance of the different approaches. We first focus on spring phenology, where we show that our deep learning model largely outperforms traditional machine learning models by $1.1$ day RMSE and $14\%$ R2 when predicting in external conditions\footnote{In a machine learning context, prediction in \emph{external condition} would be described as out-of-distribution generalisation.}, maintaining a performance level similar to the best process-based models. We then conduct similar numerical experiments for autumn phenology where the picture is more nuanced. While our deep learning model outperforms both traditional machine learning and process models when predicting in external conditions, the performance level remains pretty poor with an RMSE of $13.6$ days in line with the notorious challenge of autumn phenology modeling. In addition to these main findings, we explore different variants of our model and show for instance that it can be trained in a multi-task setting, where the joint prediction of phenology for multiple species is beneficial to its overall performance.

In summary, this study addresses the under-explored area of deep learning-based spring and autumn tree phenology, with the goal of assessing its potential for future projection compared to process models and traditional machine learning methods, ultimately contributing to enhancing the set of methodological tools available for climate change research.

\section{Materials and Methods}

\begin{figure}
    \centering
    \includegraphics[width=\linewidth]{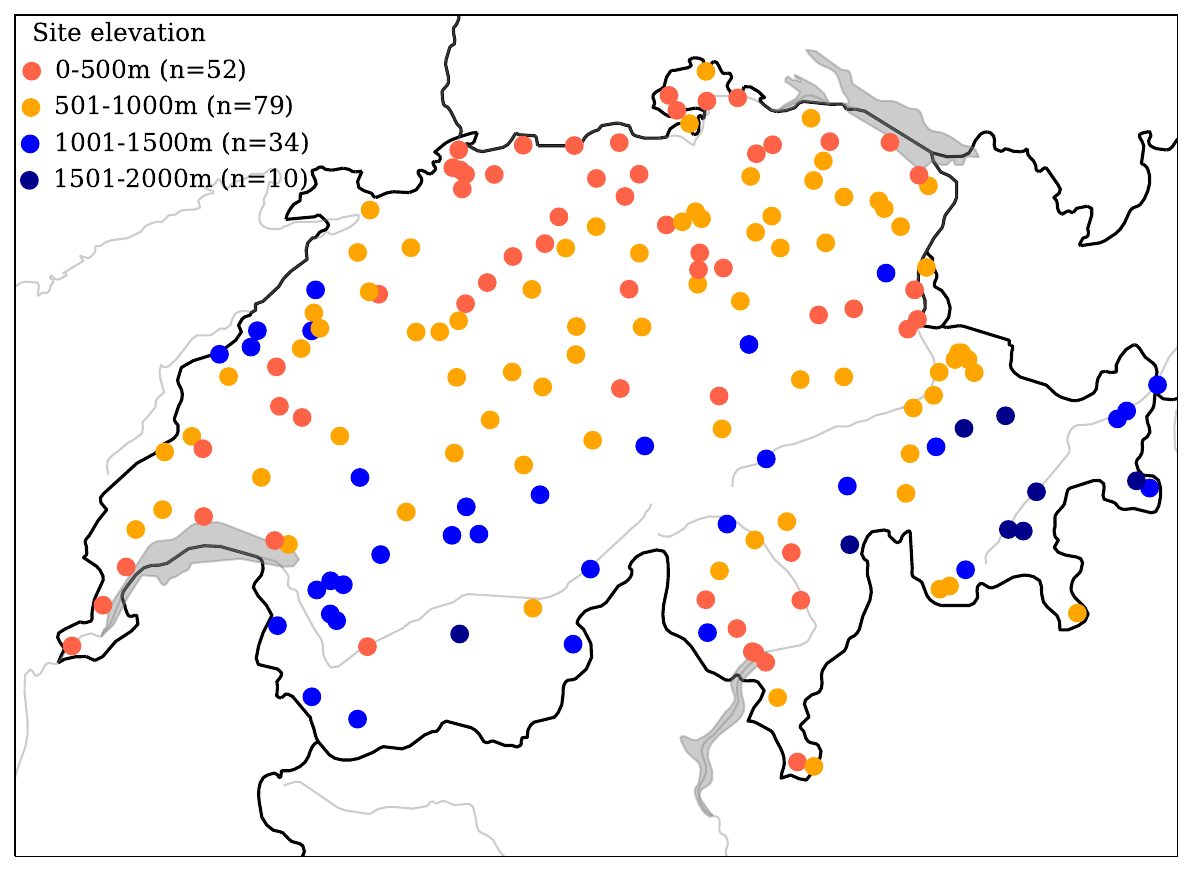}    
    \caption{Location of phenology observation sites of the Swiss Phenology Network, coloured by elevation. }
    \label{fig:spn-map}
\end{figure}

\subsection{Dataset}
We assemble a country-scale dataset for climate-phenology modeling to evaluate and compare the process-based, machine learning, and deep learning models. We construct the dataset by combining phenological observations from the Swiss Phenology Network (SPN) \citep{SPN} and daily climate time series from the DayMetCH \citep{thornton1997daymetmethods, daymetCH1, daymetCH2} dataset.

\begin{table}[h!]
  \renewcommand{\arraystretch}{0.5}

    \centering
    \caption{Species used in the current study and number of phenological observations for spring and autumn phases. Overall, our dataset comprises 67,800 phenological observations. (Observations marked with $^\dagger$ are only available since 1996) }
\begin{tabular}{llll}
\toprule
 && Spring & Autumn \\ \cmidrule{3-4}
HCH & Horse Chestnut & 5,557 & 4,990 \\
BEE & European Beech  & 6,307 & 5,903 \\
LAR & European Larch  & 7,193 & 3,265$^\dagger$ \\
EWB & European White Birch  & 3,110$^\dagger$ & 2,845$^\dagger$ \\
CRO & Common Rowan  & 3,170$^\dagger$ & 2,847$^\dagger$ \\
SLL& Small Leaved Lime  & 2,524$^\dagger$ & 2,316$^\dagger$ \\
LLL &Large Leaved Lime  & 2,633$^\dagger$ & 2,357$^\dagger$ \\
HZL & Hazel  & 6,430 & -  \\
SPR & Common Spruce  & 6,353 & -  \\ \cmidrule{3-4}
&\textbf{Total} & 43,277 & 24,523 \\  
\bottomrule
\end{tabular}
\label{tab:counts-pheno}
\end{table}

\subsubsection{Phenological records}

The SPN is a network of citizen observation of plant phenology lead by the Federal Office of Meteorology and Climatology MeteoSwiss. The SPN initiative started in 1951 enabling phenological time series spanning up to $70$ years.
The dataset of the SPN comprises 175 observation sites and covers, among other phenophases, leaf unfolding, flowering and leaf colouration. The observations are made by citizen observers following an established observational protocol \citep{defila2008spnguide}.
The observation sites are distributed across Switzerland, covering its different climatic regions, see \figref{fig:spn-map}, an elevation range of over 1700 m, from 200 to 1930 m above sea level, as well as a $2^\circ$ gradient in latitude and $4^\circ$ gradient in longitude. As a consequence of the large elevation gradient and temporal span, observations encompass a range of $1.5-14.1^\circ C$ in mean annual temperature and 550 - 4000 mm in annual precipitation.
Over the years, the number of observation sites grew and an expansion to a wider set of species was done in $1996$.
This expansion of the SPN was carried out with a continuous effort to preserve data quality and remove outliers  \citep{auchmann2018spnqa,gusewell2018representativeness, brugnara2020SPNhomo}.

\begin{figure*}   
    \begin{subfigure}[b]{0.395\textwidth}
         \centering
         \includegraphics[width=\textwidth, trim=0cm 0cm 2.45cm 0cm ,clip]{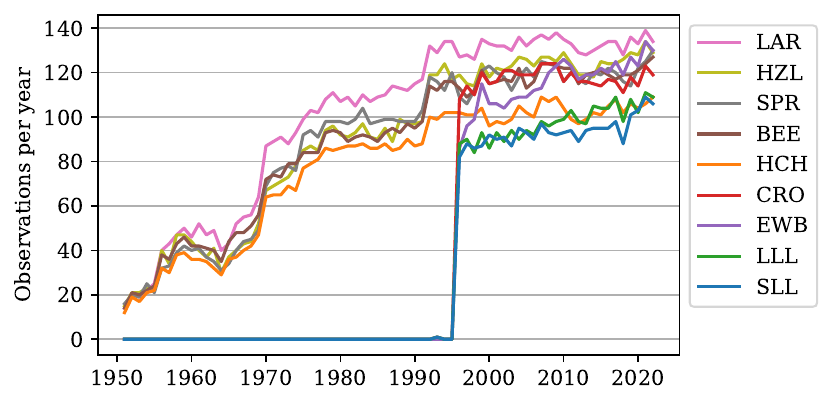}
         \caption{Leaf unfolding / Needle emergence }
         \label{fig:obs-year-spring}
    \end{subfigure}
    \hfill
    \begin{subfigure}[b]{0.4\textwidth}
         \centering
         \includegraphics[width=\textwidth, trim=0cm 0cm 0cm 0cm ,clip]{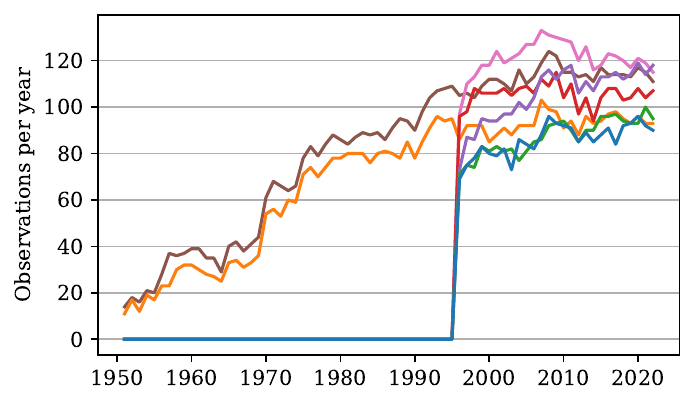}
         \caption{Leaf / Needle colouration}
         \label{fig:obs-year-spring}
    \end{subfigure}
    \hfill
    \begin{subfigure}[b]{0.15\textwidth}
         \centering
         \includegraphics[height=5cm, trim=11.6cm 0cm 0cm 0cm ,clip]{graphics/data/obs-over-time-spring.pdf}
    \end{subfigure}
    \caption{Number of phenological observations per year and per species in the Swiss Phenology Network archive for spring (a) and autumn (b). Note the addition of new species in year 1996, entailing a limitation in the available history for some species. }
    \label{fig:observations-year-spn}
\end{figure*}

In this paper, we use a subset of 9 species from this archive, broadly covering deciduous broadleaf, deciduous needleleaf, and evergreen needleleaf species. Furthermore, we focus on the prediction of leaf unfolding (LU) / needle emergence (NE) for spring and leaf (LC) / needle colouration (NC) for autumn. In some of our numerical experiments, we also use the date of flowering as auxiliary data. We select the species with the most observations, and prioritise those species with the longest available history. We show the list of species used in \tabref{tab:counts-pheno}, as well as the number of tree-year observations for each species and phenophase, amounting to a total of $67,800$ phenological observations. On \figref{fig:observations-year-spn} we show how these observations are distributed over time, highlighting the expansion of the observation program to new species that occurred in $1996$: before that year, some phenophases were not monitored in the SPN. In our selected subset of species, five out of nine (respectively two out of seven) species still have historical observations until 1950 for spring (resp. autumn) phenology.

\begin{table}[h!]
  \renewcommand{\arraystretch}{0.5}
    \caption{Climate variables used in our experiments. }
    \centering
    \footnotesize{
    \begin{tabular}{l|c}
      $T$   &  mean daily temperature ($^\circ C$)\\
      $T_{min}$ & minimum daily temperature ($^\circ C$)\\
      $T_{max}$ & maximum daily temperature ($^\circ C$)\\
      $ltm$ & long term daily mean temperature over a 30-year period ($^\circ C$)\\
      $P$ & daily precipitation sum (mm) \\
      $VPD$ & daily average water vapour pressure deficit (Pa) \\ 
      $L$ & daily photoperiod (hours) \\
    \end{tabular}
    }
    \label{tab:climate-vars}
\end{table}

\subsubsection{Meteorological time series}
We extract climate time series from the DaymetCH dataset \cite{daymetCH1, daymetCH2}. This dataset is produced using the interpolation software Daymet  \citep{thornton1997daymetmethods} to produce daily gridded maps of meteorological variables for Switzerland at a spatial resolution of $100$m. We then take the daily climate time series at the geo-location of each observation site of the SPN. We use a total of $7$ climate variables shown in \tabref{tab:climate-vars}, covering temperature, precipitation, pressure, and photoperiod.

\subsubsection{Problem statement}

\begin{figure*}
    \centering
    \includegraphics[width=.9\linewidth]{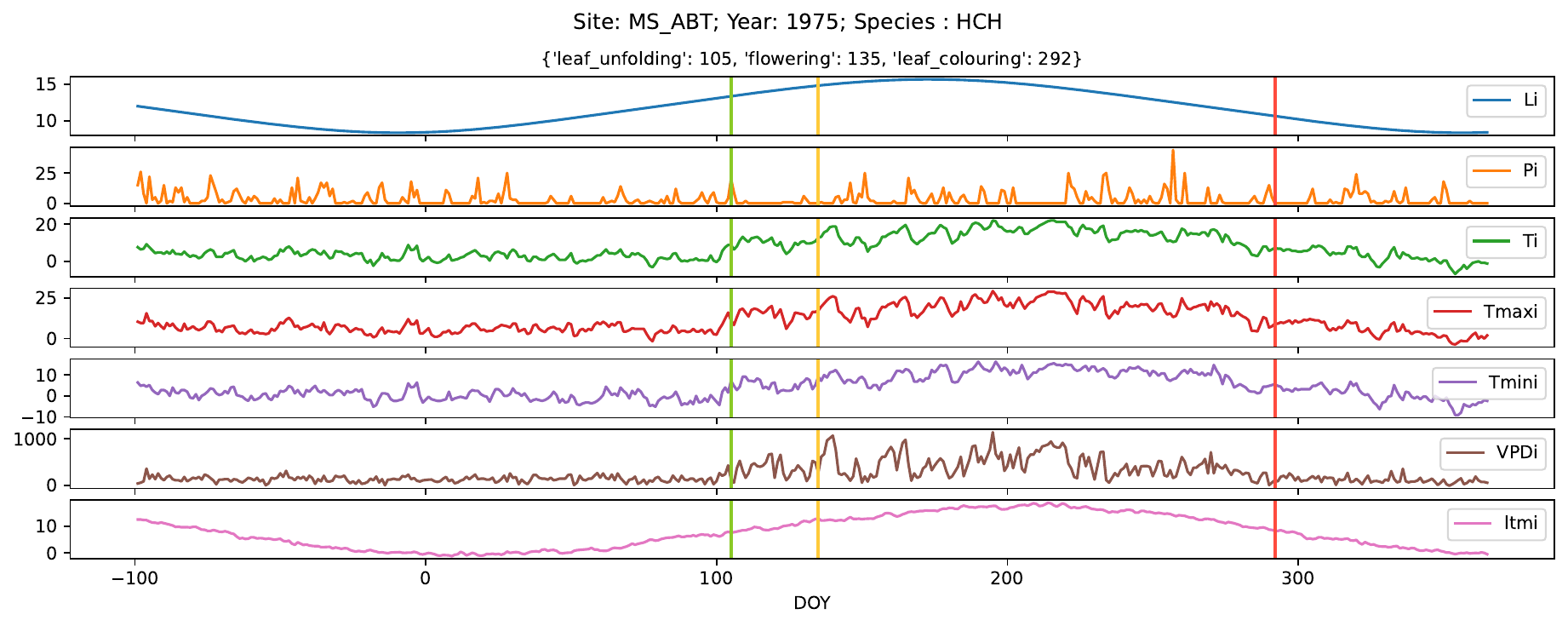}
    \caption{One sample of our dataset, the input is a 7-dimensional daily time series of climate variables for a given year and station. The target variable is the date (in day of year) of a given phenophase (coloured vertical bars).}
    \label{fig:sample-viz}
\end{figure*}

Our dataset, indexed by $n \in [ 1, N] $ is composed of pairs $(\mathbf{X}_n, Y_n)$ where $\mathbf{X}_n$ is the multi-variate daily climate time series for a specific site during a given year, and $Y_n = (y_n^1, \cdots, y_n^P)$ are the dates, expressed in day-of-year of the $P$ phenophases observed for that site-year. We show an example of such a sample on  \figref{fig:sample-viz}. The multi-variate climate time series $\mathbf{X}_n$  is of shape $S \times C $, with $C = 7$ the number of climate variables, and $S$ the number of days in the climate time series. 
For a given phenophase $p$, the task for a model $\mathcal{F}_{\theta}$ with parameters $\theta$ is to find the parameters that enable the most accurate prediction of the phenological dates based on the climate time series:

$$
argmin_{\theta} \frac{1}{N} \sum_{n} \| \mathcal{F}_{\theta}(\mathbf{X}_n ) - y_n^p\|^2 \:.
$$

\subsubsection{Distribution shifts}

\begin{figure}   
    \begin{subfigure}[b]{0.45\textwidth}
         \centering
         \includegraphics[width=\textwidth, trim=0cm .8cm 0cm 0cm ,clip]{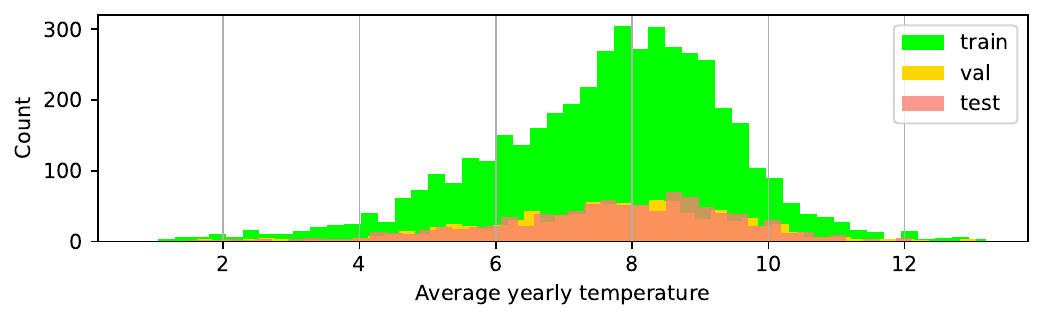}
         \caption{No shift}
         \label{fig:split1}
    \end{subfigure}
    \hfill
    \begin{subfigure}[b]{0.45\textwidth}
         \centering
         \includegraphics[width=\textwidth, trim=0cm .8cm 0cm 0cm ,clip]{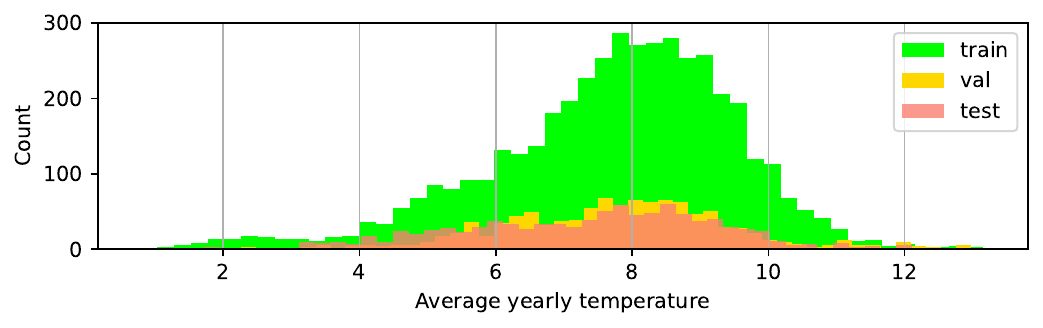}
         \caption{Random spatial split}
         \label{fig:split2}
    \end{subfigure}
    \vfill
        \begin{subfigure}[b]{0.45\textwidth}
         \centering
         \includegraphics[width=\textwidth, trim=0cm .8cm 0cm 0cm ,clip]{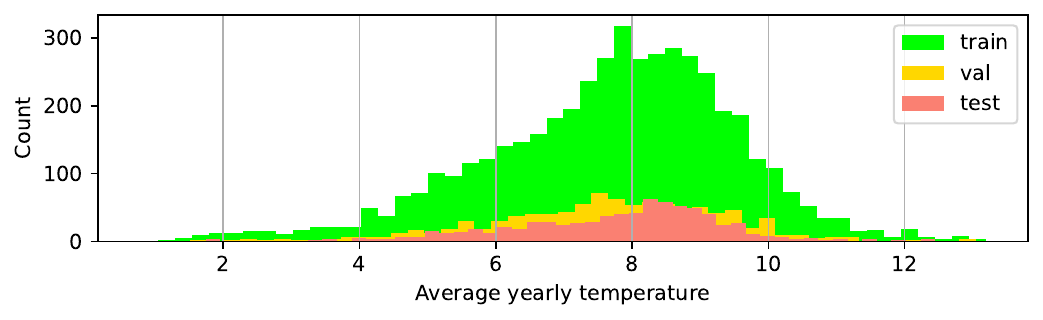}
         \caption{Random temporal split}
         \label{fig:split3}
    \end{subfigure}
    \hfill
    \begin{subfigure}[b]{0.45\textwidth}
         \centering
         \includegraphics[width=\textwidth, trim=0cm .8cm 0cm 0cm ,clip]{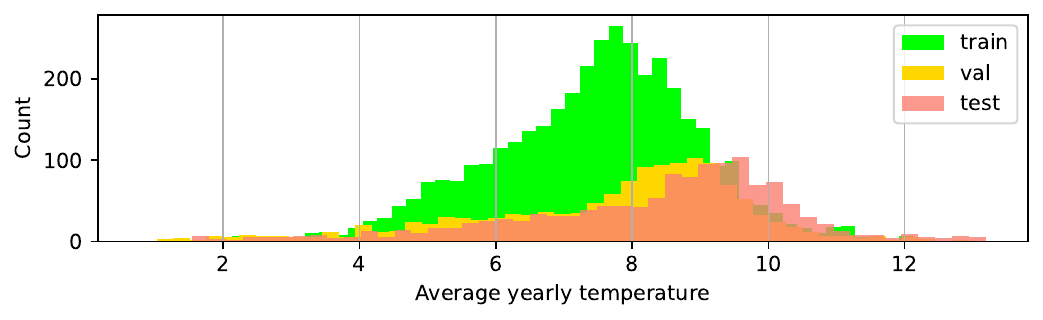}
         \caption{Structured temporal split}
         \label{fig:split4}
    \end{subfigure}
    \caption{Distribution of the average annual temperature ($^\circ C$) in the training (green), validation (orange), and test (red) sets for the four different dataset splits we investigate. Note the significant shift in climatic conditions for the structured temporal split (d).  }
    \label{fig:splits}
\end{figure}

One key aspect of the present study is to assess the robustness of the different approaches to shifts in the distribution of the data between training and testing. We propose to explore this numerically by fitting models using four different train/test splits, with different types and intensities of distribution shifts. 
While process models only require a training and testing subset, deep learning models are typically trained in combination with a \emph{validation} set. The validation set is separate from the train and test sets, and is used for intermediary evaluation during model optimisation. For each split configuration, we thus split the data by allocating each site-year $n$ into three subsets with a ratio of $70\%$ train, $15\%$ validation, and $15\%$ test.  We obtain the four different dataset splits with the following schemes:

\begin{enumerate}
    \item \textbf{No shift} Here, we randomly assign each site-year to any of the subsets. Hence, all sites and all years are likely to be represented in train, validation, and test sets. As a results, the data distributions of the three subsets are very similar as shown on Fig. \ref{fig:split1}.
    \item \textbf{Random spatial split} In this setting, we randomly split the dataset based on the observation sites: all years from a given site are allocated to either of the three subsets. 
    \item \textbf{Random temporal split} In this setting, we randomly split the dataset based on the year: all site observation of a given year are allocated to train, validation, or test. Hence, some years are only seen during training, and others only at test time.  
    \item \textbf{Structured temporal split} Lastly, we design a structured temporal split, by assigning all observations between $1951$ and $2002$ to train, $2003 \rightarrow 2012 $ to validation, and $2013 \rightarrow 2022 $ to the test set. As shown on Fig. \ref{fig:split4} the structured temporal split is the one entailing the most intense shift in distribution between training and testing, and is thus the closest to the use case of applying models for future projections under climate change scenarios. 
\end{enumerate}

We also show the magnitude of the shifts for each variable in \tabref{tab:shift} where we compute the difference between the average annual mean of each variable on the test and train set. There we can see that for the structure temporal split (4), the shift in mean annual temperature is of almost $1^\circ C$ and that the shift is four days earlier in spring and later in autumn. We will thus use the performance achieved by models on this split to assess their skill for future projection.

\begin{table}[h!]
\renewcommand{\arraystretch}{0.5}
\caption{Difference between the annual average of each variable between training and testing site-years for each dataset split configuration. 
}
\label{tab:shift}
\centering
\footnotesize{
\begin{tabular}{lrrrr}
\toprule
& \multirow{2}{*}{1.No shift}& 2.Random & 3.Random & 4.Structured \\ 
&  &  spatial &  temporal &   temporal  \\ \cmidrule{2-5}
Tmin ($^\circ C$)         &   0.01 &  -0.01 & -0.07 &   0.56 \\
Tmax ($^\circ C$)         &  -0.00 &  -0.14 & -0.04 &   1.00 \\
T ($^\circ C$)            &   0.00 &  -0.06 & -0.07 &   0.91 \\
ltm ($^\circ C$)          &   0.03 &  -0.09 & -0.19 &   0.32 \\
VPD (Pa)                  &  -0.01 &  -4.89 & -2.53 &  41.72 \\
P (mm)                    &  -0.01 &   0.02 & -0.03 &  -0.32 \\
spring (days)             &   0.0  &  -0.3  &  0.4  &  -4.0  \\
autumn (days)             &   0.2  &  -0.1  & -0.6  &   3.9  \\
\bottomrule
\end{tabular}
}
\end{table}

\clearpage
\subsection{Statistical methods}

In this section, we describe the learning-based methods we train on our dataset. We first introduce PhenoFormer, our attention-based deep learning architecture for phenology prediction. Next, we present the traditional machine learning baselines we compare it with.

\subsubsection{PhenoFormer deep learning architecture}
\label{sec:meth:dl}
We design a deep learning architecture, dubbed PhenoFormer, based on the self-attention mechanism \citep{vaswani2017attention}, a type of neural network  particularly well suited for temporal data processing. 

\begin{figure}
    \hspace{1cm}
    \includegraphics[width=1.4\linewidth, trim=0cm 8cm 26cm 0cm, clip]{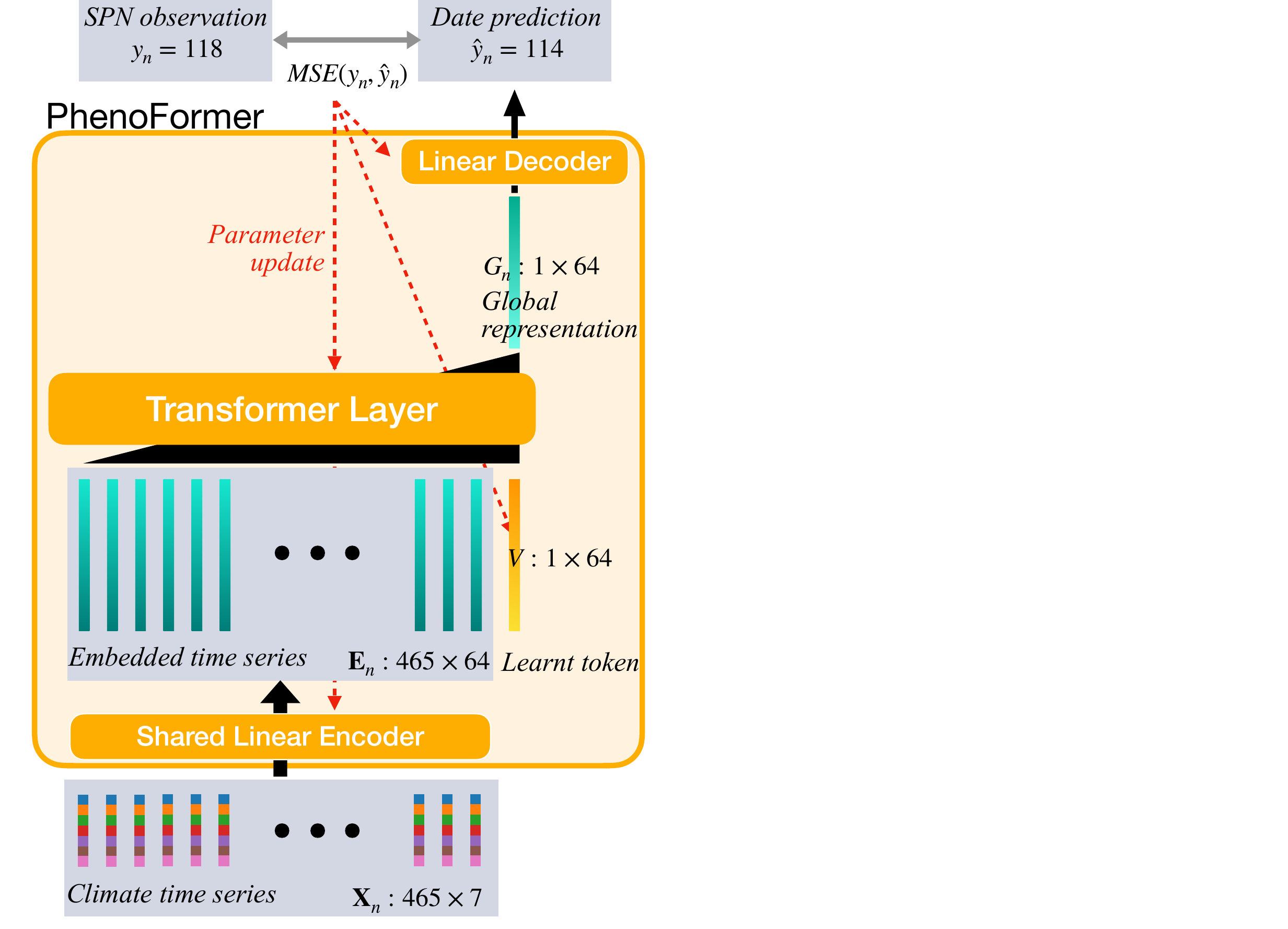}
    \caption{\textbf{PhenoFormer.} The input climate time series $\mathbf{X}_n$ is first embedded into a higher dimensional space with a shared linear layer that produces a sequence of vectors $\mathbf{E}_n$. We append a learnt token $V$ to the sequence, and process it with a transformer layer. We use the output of the transformer for the learnt token as global representation of the climate time series. The global representation is decoded into the predicted date with a linear layer. The MSE loss computed between the predicted dates and the SPN observation is then used to update the trainable parameters of PhenoFormer with gradient descent.  }
    \label{fig:archi}
\end{figure}

\textbf{Network structure} The overall architecture of our network, shown in \figref{fig:archi}, is composed of a  shared linear encoder, a Transformer layer with learnt token, and linear decoder. The network takes a climate time series as input $\mathbf{X}_n$ and outputs a predicted phenological date $\hat{y}_n$ (here we consider a single phenophase and drop the $p$ exponent). Across our experiments, we use all the days of the year at hand as well as the $100$ last days of the previous years, hence $S = 465$. 
Our network operates as follows:
\begin{itemize}
    \item \textit{Shared linear encoder:} The input climate time series $\mathbf{X}_n$ of shape $S \times C$ is first embedded into a higher dimensional space of dimension $D=64$ by a linear layer $C \rightarrow D$ ($1\times1$ convolution), applied in parallel across the time series to each input climate vector. This results in a sequence of vectors $\mathbf{E}_n : S \times D $, with  $D$ a hyperparameter of the model. 
    \item \textit{Learnt token:} We append a learnt token $V$ to the embedded sequence. $V$ is a vector of shape $D=64$ the coefficients of which are trainable parameters of the model.  
  \item\textit{Transformer layer:} We then use a Transformer encoder layer to extract temporal climatic features. 
    The Transformer layer takes as input the 
    sequence of vectors $\mathbf{E}_n$ and uses the learnt token $V$ to produce a global representation $G_n$ of shape $D$, that combines information from the whole input sequence. 
    We follow the same transformer layer design and positional encoding scheme as in the original transformer model \citep{vaswani2017attention}.
    \item \textit{Linear decoder:} Lastly, we decode the global representation with a linear layer $D \rightarrow 1 $ that returns the predicted phenophase date $\hat{y}_n$.
\end{itemize}

\textbf{Model training} The trainable parameters of our model are contained in the linear encoder, the transformer layer, the learnt token, and the linear decoder. In its default configuration with $D=64$, PhenoFormer has $\sim35$k trainable parameters. We optimise these parameters via gradient descent with Mean Squared Error (MSE) loss between the predicted date $\hat{y}_n$ and the true phenological date $y_n$ taken from the SPN data:
$$
\mathcal{L}_{MSE} =  \frac{1}{N} \: \mathlarger{\sum_{n}}  \: \| y_n - \hat{y}_n \|^2 \:\: . 
$$

\begin{figure}[]
    \hspace{2cm}
    \includegraphics[width=1.4\linewidth, trim=0cm 8cm 26cm 0cm, clip]{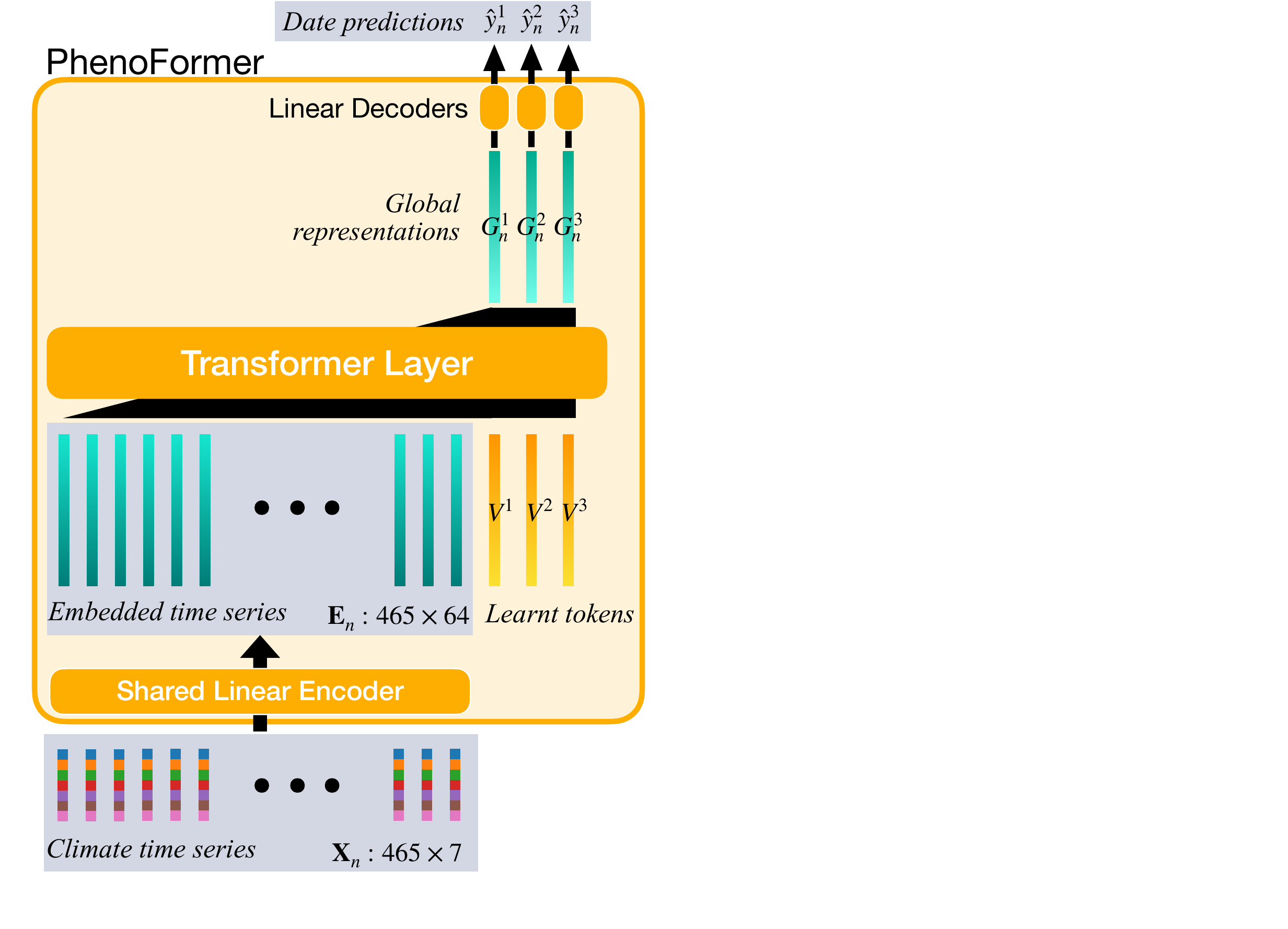}
 \caption{\textbf{Multi-task PhenoFormer.} PhenoFormer can predict multiple phenophases (different species and/or seasons) from the same input time series. We simply add a learnt token $V^p$ per phenophase. This enables the Transformer layer to produce a different global representation for each phenophase $G_n^p$. Each of them is then processed by a separate linear decoder to return the final prediction.  
    }
    \label{fig:archi-multi}
\end{figure}

\textbf{Multi-task setting} Our architecture can readily be used in a multi-task setting where the same model is used to predict different phenophases from the same input climate time series. For example, this enables the prediction of the phenological dates of all species present in a given site-year in a single model pass. We argue that this can be beneficial because the same temporal feature extractor can be relevant for different species. This also allows to pool the samples of different species to train the same model, thus increasing the amount of training data. To address the multi-task setting with PhenoFormer, we simply add one learnt token $V^p$ and one linear decoder per predicted phenophase $p$. As depicted in \figref{fig:archi-multi}, the Transformer layer now returns one global representation per phenophase $G_n^p$, which is decoded into a date prediction $\hat{y}_n^p$ by the corresponding linear decoder. 
The multi-task model is then trained by optimising the sum of the MSE losses of each individual phenophase that is predicted. Our multi-task architecture design based on different learnt tokens allows to use the same Transformer layer for different tasks, while still producing a different set of temporal climatic features for each task, and at a marginal cost in additional parameters. We show in the experiments how this leads to more robust models. In particular, we explore two settings of this multi-task approach:
\begin{itemize}
    \item \textit{multi-species:} In this setting, one single model predicts the spring or autumn phenological date of all species of our dataset. This way, we train 1 model instead of 9 (resp. 7) for spring (resp. autumn) phenology.
    \item \textit{multi-species}$+$\usym{1F338} : Here, we extend the previous setting by predicting both leaf unfolding and flowering dates for all species of our dataset. 
\end{itemize}

\textbf{Incorporating site-specific information} It can be beneficial to include additional site--specific information to improve the predictions of our model. For instance, elevation gradient has been shown to lead to genetic adaptation in tree phenology  \citep{vitasse2013elevational}. The SPN dataset provides the elevation and geo-location of each observation site. We also train a variant of our model (denoted by \usym{2609} in the result tables) to which we provide this additional information as input. To do this, we simply standardise these values and concatenate them to the climate vectors of the input time series so that the input dimension is then $C=10$, with the elevation, latitude, and longitude as constant $S \times 1$ vectors.

\textbf{Training and implementation details} We standardise each climate variable of  the input time series by subtracting its global mean and dividing by its standard deviation. 
We also standardise the phenological dates as we found this leads to more stable optimisation. 
We train our model with MSE loss and the Adam optimiser  \citep{kingma2014adam}, set with learning rate $lr=10^{-4}$, $\beta = (0.9, 0.999)$, and a batch size of $16$ site-years. 
We run a maximum of $300$ epochs and select the best epoch based on the validation set performance, and report the performance of that epoch's weights on the held-out test set.
We implement our model in Pytorch and make the code available at \url{link-upon-publication }.

\subsubsection{Traditional ML methods}
We compare our deep learning architecture to traditional machine learning and one simple statistical approach. 

We select Random Forest regressor  (RF) \citep{breiman2001RF}  and Gradient Boosted Machines (GBM) \citep{friedman2001GBM}  to represent that class of traditional machine learning approaches as they have been found to be among the most competitive in previous literature. Several studies train such models only on monthly or seasonal averages instead of daily values  \citep{dai2019gbmchina, lee2022RFKoreaColouration}. 
We thus train two variants of each model: one that takes the daily climate time series as input, and one that takes monthly averaged values. 
We use the python scikit-learn \citep{pedregosa2011scikit} implementation of those algorithms, with $n_{estimators}$ set as the square root of the total number of input features if it is above $100$ and $n_{estimators}=100$ otherwise. 
For spring phenology we also include a simple statistical model which relates average March - May temperatures to leaf emergence with a linear equation.

\subsection{Process Models}

To compare process model-based approaches to statistical models,  we select a variety of leaf emergence and colouration process models which span a range in complexity and have been found to perform well in previous process model studies  \citep{basler2016evaluating, hufkens2018phenor, spafford2023climate}. 

\subsubsection{Spring phenology}
There are three main dormancy stages simulated in process models \citep{delpierre2016temperate}. The stage of dormancy induction occurs in the autumn, when shortening photoperiods and cooling temperatures prompt preparation for winter. This is followed by endodormancy, a state regulated internally within the bud and eventually completed by sufficient exposure to chilling temperatures. Lastly, buds enter ecodormancy, which is completed once external environmental conditions, mainly temperature and photoperiod, are suitable for leaf emergence.
Our selected leaf emergence process models simulate either 1) ecodormancy release, 2) endodormancy and ecodormancy release, or 3) preceding dormancy induction, endodormancy and ecodormancy release, for a total of 8 leaf emergence process models shown in \tabref{tab:spring-pheno-process-models}. Each process model follows a similar general form: following an accumulation start date, each model simulates the accumulation of driver influence(s) until a critical threshold is reached, resulting in immediate leaf emergence.  To implement the models, we use the phenor package in R version 1.3.2  \citep{hufkens2018phenor}. For more details regarding model equations, refer to the references in \tabref{tab:spring-pheno-process-models} or the phenor package.

\subsubsection{Autumn phenology}

Our selected leaf colouration process models also span a range in complexity. Generally, these models include an accumulation of cooling temperatures and shortening photoperiod until a critical threshold is reached and leaf colouration occurs, except for the White Model (WM). Within the WM, leaf colouration occurs instantly based on a two-part triggering mechanism involving photoperiod and temperature thresholds. Five of these models simulate dormancy induction alone, while two include the influence of preceding spring leaf emergence. The Photosynthesis-Influenced Autumn with Growing Season Index (PIAG) model additionally includes the influence of estimated productivity on the timing of leaf colouration, assuming a more productive growing season corresponds to earlier leaf colouration \citep{zani2020increased}. Most models include thermal and photoperiod drivers of phenology, while the Delpierre Sigmoidal with Precipitation Influence (DMP) model also includes the influence of precipitation. To implement the models, we use the phenor package in R version 1.3.2 \citep{hufkens2018phenor} and add model functions for those which were not available based on the model equations provided in \citet{meier2023process, zani2020increased}. For more details regarding model equations, refer to the references in \tabref{tab:fall-pheno-process-models}.

\subsubsection{Process Model Calibration}

To calibrate both leaf emergence and colouration process models, we apply general simulated annealing through the pr\_fit\_parameters() function in the R phenor package, which invokes the GenSA optimization function from the GenSA package in R  \citep{xiang2013generalized}. General simulated annealing is an optimization algorithm that searches for model parameter values while converging towards a global minimum in model prediction error, similar to the process of metal cooling  \citep{chuine1998fitting}. The GenSA algorithm allows for parameters with suboptimal error to be accepted as new points for parameter assessment, thereby helping to avoid confinement in local minima  \citep{meier2023process}. In this study, calibrations are configured to seek the global root mean squared error (RMSE) between modelled and observed phenology. Each annealing begins with a starting temperature of 10,000 and a maximum of 40,000 model calls, following Hufkens et al. (2018). The parameter ranges for each model calibration are established based on previous experimental and modeling studies  \citep{basler2016evaluating,hufkens2018phenor, liu2020modeling, zani2020increased}. For the leaf emergence models, we use the default parameter ranges from the phenor package, with two exceptions: we set the minimum base temperature for forcing accumulation to 0$^\circ C$, and the earliest start date for forcing accumulation to January 1st to ensure biologically relevant forcing parameters \citep{chuine2003plant, chuine2017process, delpierre2016temperate}. For the leaf colouration models, we use the same parameter ranges presented in both Zani et al. (2020) and Meier and Bigler (2023), with two exceptions: daylength threshold parameters ranges from 8-16 hours, and cooling base temperature thresholds ranges from 10-30$^\circ C$ to reflect ranges in our meteorology data (no daylengths outside this range nor maximum temperatures greater than 30$^\circ C$).

\subsection{Evaluation}

We measure the performance of the different models with their Root Mean Square Error (RMSE) and coefficient of determination (R2) on the test set. 

$$RMSE = \sqrt{\frac{1}{N} \sum_n (\hat{y}_n - y_n)^2} \: ,$$

$$R2 = 1 - \frac{\sum_n (\hat{y}_n - y_n)^2}{\sum_n (\hat{y}_n - \mu)^2}, \quad \text{with} \quad  \mu = \text{mean}(\{y_n\}) .$$

Our numerical experiments cover a large number of  configurations with different models, and different ways to split the dataset. To consolidate our results and reduce the impact of noise we report the average performance over $10$ runs for each configuration. We use a different random seed for each of the $10$ runs, and for the splitting strategies $1,2,$ and $3$ each run also has a different site-year split.


\section{Results and Discussion}
\subsection{Spring phenology}
We first compare the performance of the different types of models in the four distribution shift configurations, and then provide more analysis of the performance of each approach individually.

\begin{table*}[ht!]
  \renewcommand{\arraystretch}{0.7}
    \centering
    \caption{
    \textbf{Spring phenology results.} Aggregated prediction performance of the models for \textit{Leaf Unfolding / Needle Emergence} on the four different dataset splits. The performances are averaged over 10 runs  and further averaged across the 9 species used in the dataset.  Best result is highlighted in bold, and second best in italic.
    }
    \vspace{.2cm}
    \begin{tabular}{lcccccccccccc}
    \toprule
    Distribution shift $\rightarrow$ && \multicolumn{2}{c}{1. None} & & \multicolumn{2}{c}{2. Rdm. Spatial} &&  \multicolumn{2}{c}{3. Rdm. Temporal } &  &\multicolumn{2}{c}{4. Struct.  Temporal} \\ 
    Model $\downarrow$ && R2 & RMSE && R2 & RMSE && R2 & RMSE && R2 & RMSE \\ \cmidrule{1-1} \cmidrule{3-4} \cmidrule{6-7} \cmidrule{9-10} \cmidrule{12-13}\\
   Process Models \\ \cmidrule{1-1}
    M1 &  & 0.52 & 9.4 &  & 0.49 & 9.5 &  & \textit{0.51} & 9.3 &  & \textbf{0.52} & \textbf{8.9} \\
    PTTs &  & 0.52 & 9.4 &  & 0.49 & 9.5 &  & 0.50 & 9.4 &  & \textbf{0.52} & \textit{9.0} \\
    PA &  & 0.51 & 9.4 &  & 0.48 & 9.6 &  & 0.49 & 9.5 &  & 0.51 & 9.1 \\
    PTT &  & 0.51 & 9.5 &  & 0.48 & 9.6 &  & 0.49 & 9.5 &  & 0.51 & 9.1 \\
    AT &  & 0.51 & 9.5 &  & 0.48 & 9.6 &  & 0.50 & 9.5 &  & 0.51 & \textit{9.0} \\
    TT &  & 0.49 & 9.6 &  & 0.46 & 9.8 &  & 0.47 & 9.7 &  & 0.50 & 9.2 \\
    SQ &  & 0.48 & 9.7 &  & 0.45 & 9.8 &  & 0.45 & 9.8 &  & 0.46 & 9.5 \\
    DP &  & 0.43 & 10.2 &  & 0.40 & 10.3 &  & 0.40 & 10.3 &  & 0.41 & 9.9 \\ \\
    PhenoFormer \\ \cmidrule{1-1}
    (e) multi-species $+ \usym{2609}+\usym{1F338}$ &  & \textbf{0.60} & \textbf{8.5 }&  & \textbf{0.51} & \textbf{9.1 }&  & \textbf{0.53} &\textbf{8.9} &  & \textit{0.51} & \textit{9.0} \\
    (d) multi-species $+ \usym{2609} $ &   & \textit{0.59} & \textit{8.6} &   & \textit{0.50} & \textit{9.2} &   &\textbf{0.53} & \textbf{8.9} &   & 0.50 & {9.1} \\
    (c) multi-species $+$\usym{1F338} &  & 0.57 & 8.9 &  & \textit{0.50} & \textit{9.2} &  & \textit{0.51} & {9.2} &  & \textit{0.51} &\textit{9.0} \\
    (b) multi-species &  & 0.56 & 8.9&  & \textit{0.50} & \textit{9.2} &  & 0.50 & {9.2} &  & \textit{0.51} & \textit{9.0} \\
    (a) single species &  & 0.56 & 9.0 &  & \textit{0.50} & \textit{9.2} &  & 0.49 & 9.3 &  & 0.48 & 9.3 \\
    \\
    Traditional ML \\ \cmidrule{1-1}
    GBM monthly &   & 0.58 & 8.7 &   & 0.48 & 9.3 &   & \textbf{0.53} & \textit{9.0} &   & 0.38 & 10.1 \\
    RF monthly &   & 0.56 & 8.9 &   & 0.49 & \textit{9.2} &   & \textit{0.51} & 9.2 &   & 0.37 & 10.2 \\
    GBM daily &   & 0.57 & 8.8 &   & \textit{0.50} & \textit{9.2} &   & 0.42 & 9.9 &   & 0.34 & 10.5 \\
    RF daily &  & 0.55 & 9.0 &  & 0.49 & 9.3  &  & 0.42 & 9.9 &  & 0.33 & 10.5 \\
    Linear &  & 0.47 & 9.8 &  & 0.44 & 10.0 &  & 0.45 & 9.8 &  & 0.45 & 9.6 \\

    \\\midrule
     Null Model &  & 0.00 & 13.6 &  & -0.02 & 13.6 &  & -0.01 & 13.5 &  & -0.09 & 13.3 \\

    \bottomrule \\
    \end{tabular}
    
    (\protect{\usym{1F338}}: the model also predicts flowering date, \protect{\usym{2609}}: geolocation and elevation of the observation sites is given as additional input.) 
    \label{tab:main-spring}
\end{table*}


We report the performance of the different models in \tabref{tab:main-spring}. The 10-fold performance average reported there are further averaged across the $9$ species of our dataset. The bottom line of the table shows the performance of the Null model predicting the mean phenophase date of the training distribution. All process-based and learning based models outperform the Null model by a large margin of up to $5$ day RMSE. We observe that the R2 performance of the Null model reflects the intensity of the distribution shift between training and testing phenology, with the lowest value of $-0.09$ reached on the structured temporal split. 

\subsubsection{Comparison of approaches}
\textbf{Overall performance} Our PhenoFormer deep architecture outperforms both process models and traditional machine learning baselines on the three first distribution shift configurations. On the structured temporal split (4), closer to the setting of future phenology projection, PhenoFormer performs on par with the best process model while the traditional machine learning baselines show quite a degraded performance. On splits 1,2, and 3, the shift in climatic conditions is quite limited, with an average change in mean annual temperature under $0.1^\circ C$. In these conditions, the deep learning model is able to make better predictions than all process models. The improvement ranges from $0.3$ day RMSE on the random spatial split (2), to a $0.9$ day improvement on split (1) where it achieves an overall RMSE of $8.5$ days, note that those metrics are averaged across the nine species of our dataset and we show in the following paragraphs that the improvements can be even larger at species level.  The traditional machine learning baselines perform also well on these splits and only slightly underperform PhenoFormer. On the structured temporal split (4), the shift in climatic condition is more pronounced with $+0.9^\circ C$ between testing and training. In such conditions, the traditional machine learning baselines perform more poorly, while the PhenoFormer maintains a performance level on par with the best process-based models at around $9$ day RMSE.
In summary, our results show that our deep learning model presents a better trade off than traditional machine learning methods between predictive capacity, shown on splits $1\rightarrow3$, and robustness to external climatic conditions, shown on split $4$, demonstrating its potential for phenology projection under climate change scenarios. 


\textbf{Robustness to different distribution shifts} For both process models and PhenoFormer, the spatial distribution shift (2) is the most challenging and incurs the largest drop in performance compared to the split without distribution shift (1). 
Despite having qualitatively similar sensitivity to the different kinds of shifts, the deep learning model still brings an improvement over the process models of $0.3$ and $0.4$ day RMSE on the spatial (2) and random temporal (3) shift respectively. 
Conversely to learning methods, process models exhibited improved performance for the prediction of out-of-distribution site-years within the structured temporal split (4) compared to that of the non-distribution shift split, with a 0.2 to 0.5 day reduction in RMSE across models. This is likely due to a combination of factors: process models have been shown previously to be more robust to distribution shifts than statistical models  \citep{asse2020process}, possibly due to their simulation of causal relationships with physically interpretable and fewer parameters. Additionally, warmer springs have been shown to lead to more consistent leaf emergence timings  \citep{denechere2021within}, which may lead to a lower magnitude of prediction error relative to typical or colder springs. To further explore this, we ran additional process model calibrations training with 30-year moving windows from 1951 to 1980, and testing with 1) the same period at external sites and 2) the following 30-year period at external sites, such that training from 1951-1980 would mean testing from 1981-2010. For each of these tests, all leaf emergence models performed better with the later test, with an average improvement of 0.7 days RMSE.

\begin{figure*}[h!]
    \centering
    \begin{subfigure}[b]{0.4\textwidth}
         \centering
         \includegraphics[width=\textwidth]{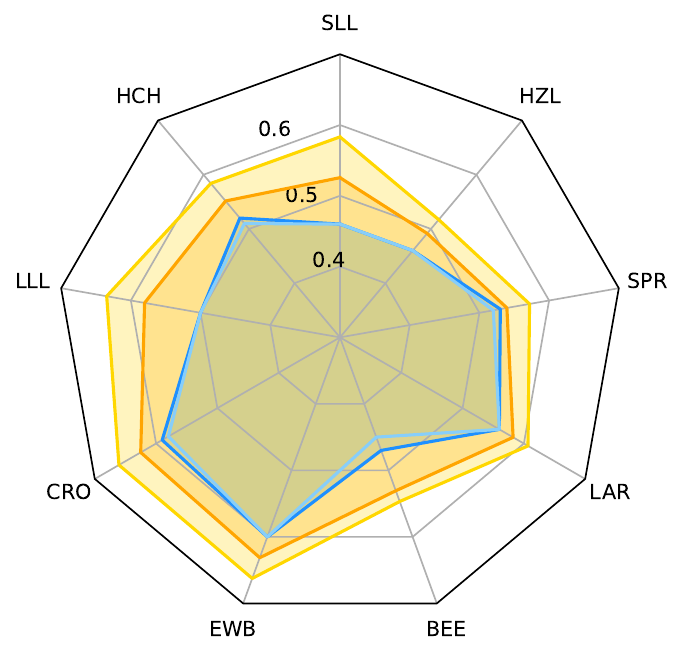}
         \caption{Split 1}
         \label{fig:radar-split1}
    \end{subfigure}
    \hfill
    \begin{subfigure}[b]{0.45\textwidth}
         \centering
         \includegraphics[width=\textwidth]{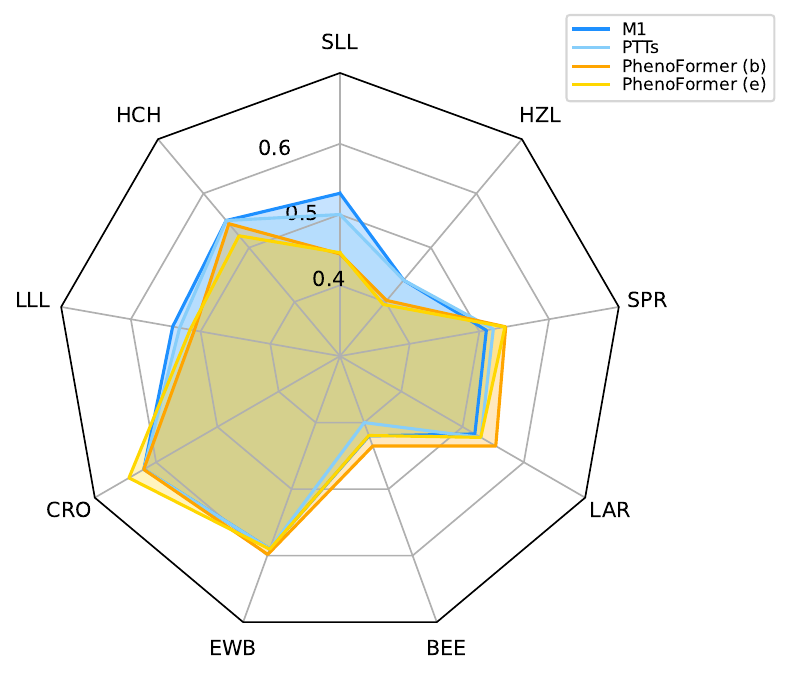}
         \caption{Split 4}
         \label{fig:radar-split4}
    \end{subfigure}
    \caption{R2 performance of the different models for each species of our dataset on splits 1 and 4. (Abbreviations: SLL: Small leaved lime, HZL: Hazel, SPR: Spruce, LAR: Larch, BEE: European Beech, EWB: European white birch, CRO: Common rowan, LLL: Large leaved lime, HCH: Horse chestnut)}
    \label{fig:radar-spring}
\end{figure*}

\textbf{Comparison per species} We show the per-species performance of two process models and two-configurations of our deep learning model on \figref{fig:radar-spring}, \tabref{tab:perspec-split1}, and \tabref{tab:perspec-split4}. On split (1) without distribution shift, the deep learning model improves the performance compared to the process models on all species, with gains ranging from $4\%$ R2 for spruce (SPR) to $13\%$ for large leaved lime (LLL). As shown on \tabref{tab:perspec-split1}, this corresponds to a reduction in RMSE of $0.5$ days for spruce and $1.4$ days for lime. On the more challenging split 4, the situation is different as the performance gaps are smaller and the deep learning model performs better on beech (BEE), larch (LAR), birch (EWB), and rowan (CRO), while the process models yield better performance on the remaining species. 
On this split, the fact that some species (highlighted by $^\dagger$ ) were only observed since $1996$ is noteworthy. For these species the shift between training and testing conditions is less marked due to the shorter history. Nevertheless the general trend of similar performance between PhenoFormer and process models remains true for both these species, and the ones with training conditions starting in $1950$.

\begin{table*}[h!]
  \renewcommand{\arraystretch}{0.5}
    \caption{Per species RMSE on split 1.}
    \centering
    \begin{tabular}{lcccccccccc}
\toprule
Model  & SLL  & HCH & LLL & CRO & EWB & BEE & LAR & SPR & HZL & Avg \\ \cmidrule{2-10}
PTTs & 8.5 & 10.0 & 8.4 & 9.4 & 8.9 & 8.1 & 9.8 & 9.9 & 11.6 & 9.4 \\
M1 & 8.5 & 10.0 & 8.4 & 9.3 & 8.9 & 7.9 & 9.8 & 9.8 & 11.6 & 9.4 \\
PhenoFormer multi-species $+ \usym{2609}+\usym{1F338}$ (e) & \textbf{7.5} & \textbf{9.1} & \textbf{7.0} & \textbf{8.3} &\textbf{ 8.0} & \textbf{7.2} & \textbf{9.2} & \textbf{9.3} & \textbf{10.9} &\textbf{ 8.5} \\
PhenoFormer multi-species (b) & 8.0 & 9.6 & 7.6 & 8.7 & 8.4 & 7.5 & 9.5 & 9.7 & 11.2 & 8.9 \\
\bottomrule
\end{tabular}
    \label{tab:perspec-split1}
\end{table*}

\begin{table*}[h!]
  \renewcommand{\arraystretch}{0.5}
    \caption{Per species RMSE on the structured temporal split 4. ($^\dagger$: species for which the observations start in 1996)}
    \centering
    \begin{tabular}{lcccccccccc}
    \toprule
Model  & SLL$^\dagger$ & HCH & LLL$^\dagger$ & CRO$^\dagger$ & EWB$^\dagger$ & BEE & LAR & SPR & HZL & Avg \\ \cmidrule{2-10}
PTTs & 8.6 & 7.9 & 8.4 & 8.8 & 9.1 & 7.8 & 9.6 & 10.2 & 10.4 & 9.0 \\
M1 & \textbf{8.3} &\textbf{ 7.8} & \textbf{8.3} & 8.8 & 9.1 & 7.7 & 9.7 & 10.2 & \textbf{10.3} & \textbf{8.9} \\
PhenoFormer multi-species $+ \usym{2609}+\usym{1F338}$ (e) & 9.0 & 8.1 & 8.5 & \textbf{8.6} & 9.1 & 7.6 & 9.6 & 10.1 & 10.6 & 9.0 \\
PhenoFormer multi-species (b) & 9.0 & 7.9 & 8.5 & 8.9 & \textbf{9.0 }& \textbf{7.5} & \textbf{9.4} & \textbf{10.0} & 10.5 & 9.0 \\ \bottomrule
\end{tabular}
    \label{tab:perspec-split4}
\end{table*}

\subsubsection{Separate analysis}

\textbf{PhenoFormer configurations} We run different configurations of our deep learning architecture.
We first compare the single species to the multi species variants. In the single species configuration one separate model is trained for each of the nine species of our dataset, similarly to process models. In the multi-species configuration we train a single model to predict the phenophase dates of all the species at once, using the multi-task architecture described in section \ref{sec:meth:dl}. On the first three splits, this multi-species model performs equally or slightly better to the single-species variant. Note that beyond the regression performance, the multi-species configuration also brings a nine-fold reduction of the computational cost at training and inference time. Extending this approach to datasets with even larger number of species can thus increase the interest of this approach. Furthermore, on the structured  temporal split (4), the multi-species model significantly improves the performance compared to the single species variant. We hypothesise that the constraint of extracting features for multiple species encourages the shared temporal encoder to learn more robust features. Overall, multi-species training yields better performance and lighter computational load so we advocate for its adoption in further exploration of deep learning based climate phenology modeling.
Next, we extend the scope of the multi-task setting by predicting leaf emergence and flowering of all species with a single model, this configuration is denoted with $\usym{1F338}$ in \tabref{tab:main-spring}, and brings a marginal improvement on some of the splits.
Last, we also train a variant of our PhenoFormer model to which we provide the elevation and geo-coordinates of the site under consideration as additional input. We denote this configuration with $\usym{2609}$ in  \tabref{tab:main-spring}. This seems to enable the model to learn some site-specific priors that are associated with a $0.3$ day decrease in RMSE on splits 1 and 3. 
Training one model per site is another way to  capture such local specificities of the individuals and environmental conditions, but comes at a higher computational cost. We argue that learning based methods enable a convenient middle  way by training a single model and providing some site specific metadata as additional input. As shown by our $\usym{2609}$ variant, this information can successfully be leveraged by the model to refine its predictions.

\textbf{Traditional ML} The results of the traditional machine learning baselines are reported in the third section of \tabref{tab:main-spring}. The GBM consistently outperforms the RF regressor. Interestingly, even if the prediction is made at a daily scale, taking monthly averaged input time series instead of daily data seems mostly beneficial to those models . One can see this as a form of feature engineering that helps the performance of traditional ML methods. We leave it for further research to explore if that results would still hold on a dataset covering a larger region and more diversity in the climate conditions. When compared to our deep learning architecture, the GBM model brings a similar improvement over the process-based models on the first three split configurations. However, on the more challenging fourth split, all traditional ML models show very poor performance compared to the deep learning and process-based models. We argue that these results show how deep learning models are better suited than traditional ML methods for out-of-distribution future projection. 

\textbf{Process-based models }
Among leaf emergence process models, both simple and complex models performed well, consistently outperforming the Null model by more than 3 days RMSE. Although the M1 model performed best across each split, the next best models were within 0.1 days RMSE, indicating that no singular process model was particularly outstanding. This is consistent with previous process model intercomparison studies including the same models though for larger regions   \citep{basler2016evaluating, hufkens2018phenor}. The two best process models, M1 and PTTs, are both simple photo-thermal ecodormancy models which include a photoperiod-modified accumulation of either growing degree days or a sigmoidal temperature response, respectively. Among the next best models were the PA and AT models, which are two-phase models including the processes of endodormancy and ecodormancy with either a parallel or alternating accumulation mechanism, respectively. The worst performing model among process models was the DP model. Despite offering a more comprehensive representation of the processes leading to leaf emergence based on experimental findings --- incorporating dormancy induction, endodormancy, and ecodormancy, with both photoperiod and temperature influences   \citep{caffarra2011modelling} --- this complex model was outperformed by simpler models. This phenomenon has also been observed in previous model intercomparison studies  \citep{basler2016evaluating, melaas2016multiscale, schadel2023using}, and may be due to greater model complexity necessitating a greater number of free parameters and a less certain calibration [and also to the relative influence of ecodormancy versus endodormancy and dormancy induction on the timing of leaf emergence].


\textbf{Attention mask visualisation} The extraction of temporal climatic features in the Transformer layer of PhenoFormer is enabled by its self-attention module. We refer the reader to  \citet{vaswani2017attention} for a detailed explanation of this method. In general terms, self-attention produces an output by computing a weighted sum over the temporal dimension of the input time series of vectors $\mathbf{E}_n$. The coefficients of this weighted sum are called \emph{attention masks}, as they explicitly show which time steps of the input sequence were most important in making the prediction. The attention masks are also data-dependent: the coefficients are computed from the input data so that each sample can yield a different attention mask. 
We plot the average attention masks of the PhenoFormer on the input climate time series for each species on \figref{fig:spring-att-viz}. We compute the average attention mask across all samples of the test set. We also show the standard deviation of the attention mask computed on the 10 runs. The dark dashed line represents a uniform repartition of attention where each date of the time series receives equal attention. The dates of the time series that are most critical to predict the phenophase date are those for which the average attention is above that threshold. 

For all species, we can see that the PhenoFormer bases its prediction mostly on dates between day of year $50$ and $150$, quite close to the date of leaf emergence (shown in green). This suggests that drivers and processes during ecodormancy are more important than earlier driver data, and that chilling is not limiting in our dataset. This observation is in line with the lower performance of process models simulating endodormancy and dormancy induction that we observed in the previous paragraph. When comparing the attention masks between the different species, we can see variations in the length and position of the important time window. 
For some species such as Horse chestnut, Small leaved lime, or European beech, the attention is concentrated on a $\sim 50$ day time window that is close to the date of leaf emergence, while for other species such as larch, hazel or spruce, the attention is spread on a wider time window of $\sim 100$ days. These variations could be hints at the relative importance of photoperiod, chilling, and forcing for different species. We leave it for further research to explore this direction in a more quantitative way.

\subsection{Autumn Phenology}

We now present the results of the different models for autumn phenology prediction in \tabref{tab:autumn-main}. We use the same four dataset splits corresponding to different distribution shifts. In addition to the previously introduced variants of our PhenoFormer model, we evaluate a variant to which we give the date of the spring phenophase date as input (f). 

\begin{table*}[th!]
  \renewcommand{\arraystretch}{0.7}
        \caption{\textbf{Autumn phenology results.} 
        Aggregated prediction performance of the models for \textit{Leaf / Needle Colouration} on the four different dataset splits. The performances are averaged over 10 runs  and further averaged across the 7 species used in the dataset. Best result is highlighted in bold. 
    }   
            \centering

    \begin{tabular}{lcccccccccccc}
    \toprule
    Distribution shift $\rightarrow$ && \multicolumn{2}{c}{1. None} & & \multicolumn{2}{c}{2. Rdm. Spatial} &&  \multicolumn{2}{c}{3. Rdm Temporal } &  &\multicolumn{2}{c}{4. Struct. Temporal } \\ 
    Model $\downarrow$ && R2 & RMSE && R2 & RMSE && R2 & RMSE && R2 & RMSE \\ \cmidrule{1-1} \cmidrule{3-4} \cmidrule{6-7} \cmidrule{9-10} \cmidrule{12-13}
    \\
   Process Models \\ \cmidrule{1-1} 
DM1Za20 &   & 0.04 & 13.6 &   & 0.02 & 13.6 &    & 0.03 & 13.4 &   & -0.06 & 13.9 \\
JM &   & 0.04 & 13.6 &   & \textbf{0.03} & \textbf{13.5} &   & 0.03 & 13.4 &   & -0.06 & 13.9 \\
PIAG &   & 0.04 & 13.6 &   & 0.02 & 13.6 &   & 0.03 & 13.4 &   & -0.07 & 14.0 \\
DPDIs &   & 0.04 & 13.6 &   & 0.03 & \textbf{13.5} &   & 0.02 & 13.4 &   & -0.07 & 14.0 \\
DPDI &   & 0.04 & 13.7 &   & 0.02 & 13.6 &  & 0.02 & 13.4 &   & -0.06 & 14.0 \\
DM1 &   & 0.04 & 13.7 &   & 0.02 & 13.6 &   & 0.02 & 13.5 &   & -0.07 & 14.0 \\
DM2 &   & 0.03 & 13.7 &   & 0.02 & 13.6 &  & 0.02 & 13.4 &   & -0.06 & 14.0 \\
DM1s &   & 0.03 & 13.7 &   & 0.01 & 13.6 &    & 0.02 & 13.5 &   & -0.08 & 14.1 \\
WM &   & 0.00 & 13.9 &   & -0.01 & 13.8 &   & -0.02 & 13.7 &   & -0.11 & 14.2 \\
DMP &   & 0.00 & 13.9 &   & -0.01 & 13.8 &  & -0.02 & 13.7 &   & -0.12 & 14.3 \\ \\
    PhenoFormer \\ \cmidrule{1-1}
(d) multi-species $+ \usym{2609}$ &   & \textbf{0.17} & \textbf{12.6} &   & 0.01 & 13.7 &   & 0.09 & 12.9 &   & -0.03 & 13.7 \\
(b) multi-species  &   & 0.11 & 13.1 &   & {0.02} & {13.6} &   & 0.01 & 13.4 &   & -0.04 & 13.8 \\
(f) single species $+$ \leafNE &   & 0.10 & 13.2 &   & 0.02 & 13.6 &   & 0.03 & 13.3  &   & \textbf{-0.01} & \textbf{13.6} \\
(a) single species &   & 0.09 & 13.2 &   & 0.02 & 13.6 &   & 0.01 & 13.4  &   & {-0.02} & {13.7} \\ \\ 
    Traditional ML \\ \cmidrule{1-1}
GBM monthly &   & \textbf{0.17} & \textbf{12.6} &   & -0.01 & 13.8 &   & \textbf{0.11} & \textbf{12.7} &   & -0.23 & 14.9 \\
RF monthly &   & 0.12 & 13.0 &   & -0.02 & 13.9 &   & 0.10 & 12.8 &   & -0.14 & 14.4 \\
GBM daily &   & 0.14 & 12.8 &   & \textbf{0.03} & 13.6 &   & 0.06 & 13.1 &   & -0.27 & 15.1 \\
RF daily &   & 0.09 & 13.2 &   & -0.00 & 13.8 &   & 0.04 & 13.2 &   & -0.18 & 14.6 \\  \\ \midrule
    Null Model &   & 0.00 & 13.9 &   & -0.01 & 13.8 &   & -0.02 & 13.7 &   & -0.09 & 14.6 \\
    \bottomrule
    \end{tabular}

( \usym{2609}: geolocation and elevation of the observation sites is given as additional input. 
    \leafNE : date of the spring phenophase is given as input)
    \label{tab:autumn-main}
\end{table*}

\textbf{Overall performance} In line with other studies, the overall performance of the autumn phenology models is much poorer than what is achieved on spring phenology. All configurations tested here achieve an RMSE that stays in the range of two weeks, with the best value of $12.6$ day RMSE achieved on split (1). 
Across methods, the improvement in RMSE compared to the Null model remains much smaller ($\sim 1$day) than what we observed for spring phenology ($\sim 5$days). Our results confirm the challenge of autumn phenology. Part of the explanation could be that autumn phenological phases are less visually distinct than spring phenology. The colouration of leaves is typically a progressive process while the appearance of buds and flowers is more marked. This can lead to a lower certainty in autumn observations with more inter-observer variability, and make the autumn problem intrinsically harder. Integrating remote sensing or phenocam imagery in the modeling pipeline could help remove part of the ambiguity. Still, there are likely other elements at play. It could be that complementary sources of data are needed to enable better predictions. Soil moisture, for example, has been found to be an important factor in previous studies  \citep{estiarte2015alteration, delpierre2017tree}.
Nevertheless, despite the overall lower level of performance, we can still observe differences in the robustness of the three types of models we evaluate. Across dataset splits, the process-based models improve over the Null model by only $0.3$ day RMSE. Concurrently, on splits (1) and (3), with limited distribution shift, PhenoFormer and the traditional ML approaches show a better performance with an improvement over the Null model of $\sim 1$ day RMSE. Lastly, on split (4) with a large distribution shift only PhenoFormer improves over the Null model by $1$ day RMSE while the traditional ML approaches perform worse than the Null model. In summary, we observe a similar pattern as for spring phenology suggesting that PhenoFormer is better suited at future autumn phenology projection than traditional machine learning algorithms.

\textbf{Process-based models}
In contrast to the leaf emergence process models, all leaf colouration models performed poorly, often only narrowly outperforming or even failing to outperform  the Null model. However, the RMSE of leaf colouration models being approximately two weeks is consistent with other process model intercomparison studies of autumn phenology  \citep{delpierre2009modelling, liu2020modeling, meier2024process}. Also unlike the leaf emergence process models, all leaf colouration process models performed worse with structured temporal splits compared to other splits. Performances between leaf colouration models were very consistent with one another, with the difference in RMSE between the best and worst model for each split being less than 0.4 days RMSE, while for leaf emergence models there was a spread of 1.0 day RMSE between the best and worst models. This  consistent poor performance between colouration models, despite different driver inclusions and representations, suggests that further work is needed to accurately capture autumn phenology mechanisms.

\textbf{Spring phenophase as input} Variant (f) of our PhenoFormer architecture predicts the colouration date for a single species based on the input climate time series \emph{and} the date of the spring phenophase for that particular tree-year. This variant achieves an overall similar performance to variant (a) that does not have access to the spring phenophase date. Hence, this additional information does not seem helpful for better autumn phenology prediction. 
Likewise, among colouration process models, those including the estimated timing of leaf emergence (DM1s and DPDIs) did not outperform other process models. Liu et al. (2020) found that the inclusion of carryover leaf emergence effects improved the prediction of leaf senescence for a given site over time for some species, but was less beneficial for predicting between sites. Liu et al. (2020) suggested this limited benefit may be due to the representation style used in the DM1s and DPDIs models, with leaf emergence affecting leaf colouration thresholds in a linear fashion. In reality there is a complex combination of influences related to leaf life span, soil water availability, frost risk, early summer growth and seasonal photosynthetic capacity \citep{liu2020modeling, zani2020increased, zohner2023effect}. 
This suggests that additional representations, cues, and site-level characteristics need to be investigated for the simulation of the influence of leaf emergence timings on leaf colouration with either machine learning or process models.

\begin{figure*}
    \centering
    \includegraphics[width=\linewidth]{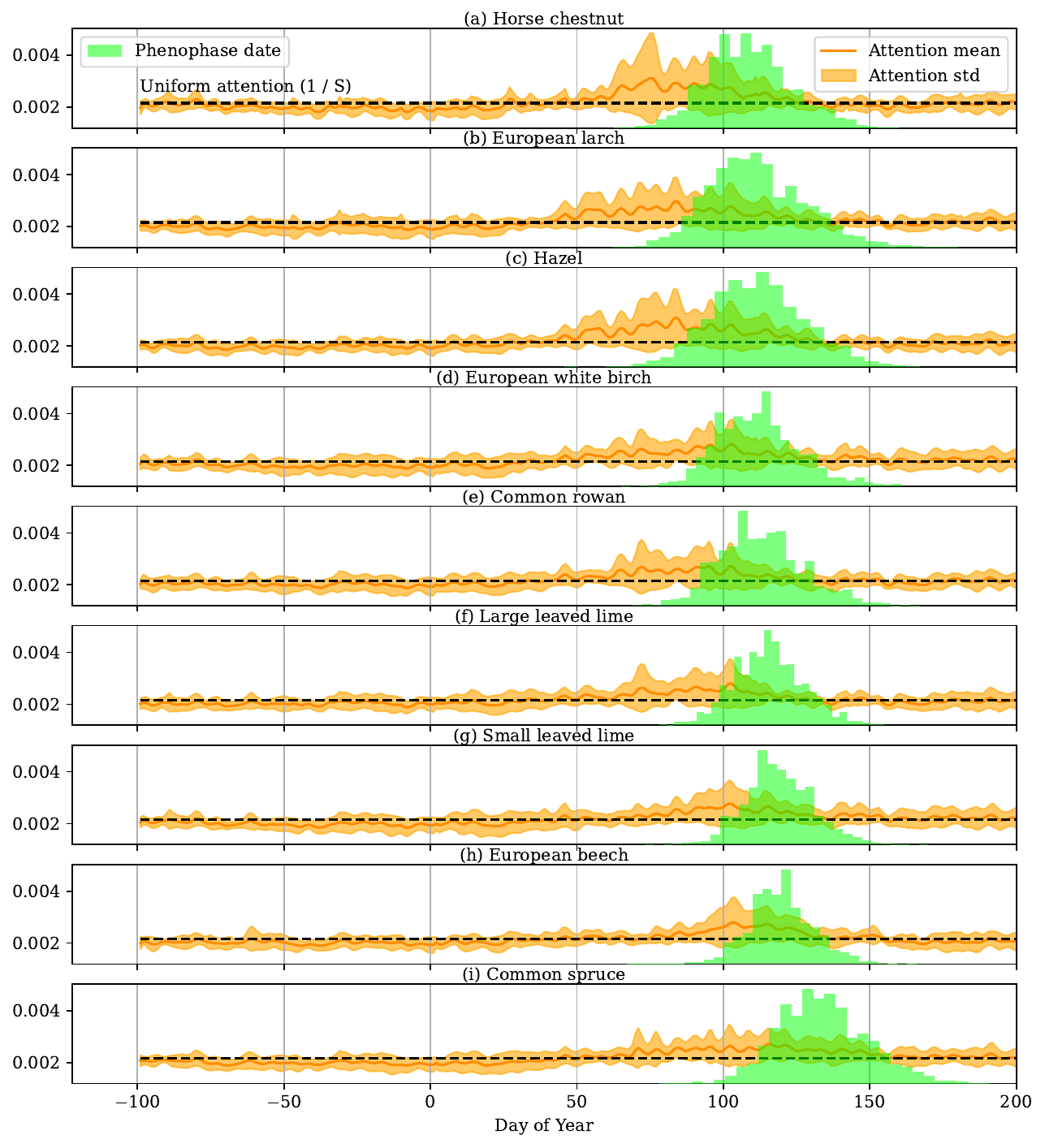}
    \caption{Average attention of the spring single-species PhenoFormer model (orange line). We compute the average across all samples of the test set and across all attention heads. We show one standard deviation of this average attention (shaded orange area) computed across the 10 runs, and the dashed black line gives the magnitude of uniform attention where each date has the same importance. The dates for which the average attention is above the uniform attention is above the uniform threshold indicate the time period of the input climate time series that was most important for the prediction.  }
    \label{fig:spring-att-viz}
\end{figure*}

\begin{figure*}
    \centering
    \includegraphics[width=.8\linewidth]{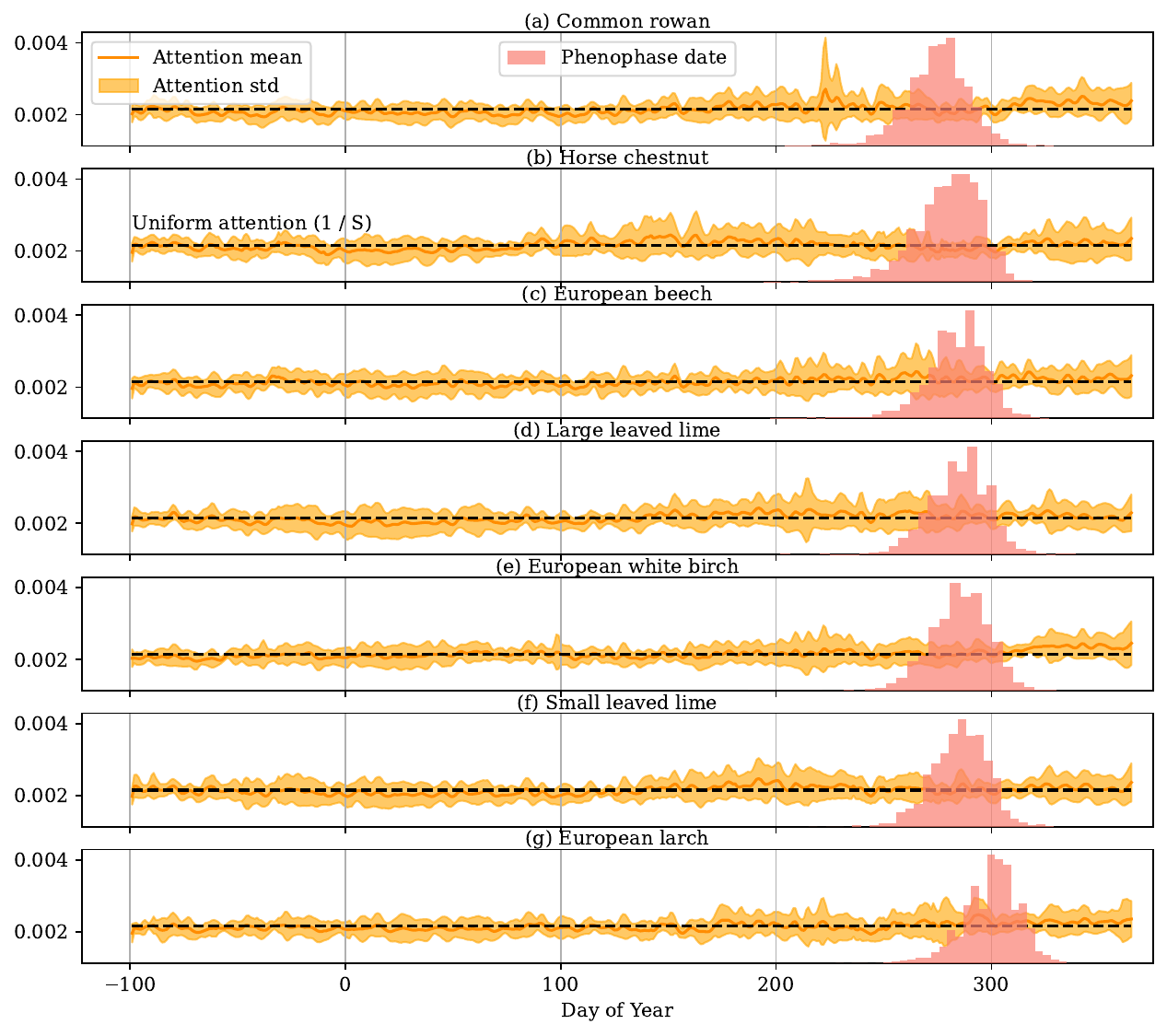}
    \caption{Average attention of the autumn single-species PhenoFormer model (orange line). We compute the average across all samples of the test set and across all attention heads. We show one standard deviation of this average attention (shaded orange area) computed across the 10 runs, and the dashed black line gives the magnitude of uniform attention where each date has the same importance. The dates for which the average attention is above the uniform attention is above the uniform threshold indicate the time period of the input climate time series that was most important for the prediction.  }
    \label{fig:autumn-att-viz}
\end{figure*}

\textbf{Attention visualisation} We show the average attention of the single-species PhenoFormer model for autumn on \figref{fig:autumn-att-viz}. Compared to the attention masks of spring phenology (\figref{fig:spring-att-viz}), the attention masks for autumn show less clear trends, with an average attention that remains very close to uniform attention. This suggest that other variables might be necessary to better capture the dynamics of autumn phenology. Yet, though less pronounced, we still note some patterns of higher attention in the months before senescence.

\clearpage
\section*{Conclusion}
In this article we introduced a novel deep learning architecture for tree phenology modeling dubbed PhenoFormer. We used a country-scale dataset of climate time series and phenological observations of nine species for spring and autumn containing around $70,000$ observations over a $70$-year period. To investigate the potential of deep learning based approaches for phenology modeling, we compared the performance of PhenoFormer to a large set of established process-based models as well as three traditional statistical approaches. Of particular interest in our experiments was the robustness to changes in the data distribution between training and testing conditions. 

Our deep learning modeling is better suited than traditional ML to tackle the challenge of phenology prediction under significant climatic conditions shift. For both spring and autumn phenology, the RF and GBM models were largely outperformed on the structured temporal split (4), operating with a $+0.9^\circ$C  change in mean annual temperature and a $4$ day shift in average phenology between  training and testing conditions. On this split, the GBM model achieves a $10.1$ day RMSE averaged across species for spring phenology, performing worse than a simple linear model, and a $14.9$ day RMSE for autumn phenology, performing worse than the null model. In comparison, our deep learning model maintained a competitive performance on this split with a $9.0$ day RMSE on spring and $13.7$ day RMSE on autumn phenology.  We believe that the superior temporal feature extraction capacity of attention-based models such as the PhenoFormer enables this greater robustness. Furthermore, we established in our experiments that deep learning also allows easy integration of additional metadata and multi-tasking over multiple species which further help the performance and robustness to shifts. 

In comparison to process-based models, our experiments with PhenoFormer present a more nuanced picture. In settings with no or limited shift between training and testing conditions, our deep learning model brings significant improvements over the best process-based model with a reduction of up to $1.4$ day RMSE on spring phenology. These performance improvements can prove useful, for instance, for outlier detection in citizen observation initiatives such as the Swiss Phenology Network, or for spatial interpolation of phenological observations. When predicting under significant climatic condition shift, PhenoFormer performs on par with the best process-based models for spring phenology at $9.0$ day RMSE and slightly better than the best process-based model on autumn phenology at $13.7$ day RMSE. The fact that deep learning approaches are challenged by such shifts is not surprising, as machine learning is based on the hypothesis of identically distributed samples at train and test time. In this regard, we argue it is already an encouraging result that PhenoFormer maintains a performance level comparable to the best process-based approaches, without any specific methodological effort to address the distribution shift. Out-of-distribution generalisation is a very active area of research in machine learning \citep{liu2021OOD}, and further research should explore how to leverage such techniques to enable more robust phenological projections under climate change scenarios. Another direction for further exploration is the integration of process-based and deep learning models into hybrid modeling approaches. Process-based models have some built-in robustness thanks to their simplicity, and some models such as the M1 model were explicitly designed for better robustness to climatic shifts. On the other hand, parameter-rich deep learning models can extract complex temporal patterns from climatic time series. Combining the strengths of both approaches in hybrid models can be another pathway to robust phenology projection. 

In conclusion, we believe that our first inquiry into deep learning-based tree phenology modeling makes a strong case for its potential. In out-of-distribution settings PhenoFormer outperformed traditional ML methods and performed competitively to the best process-based models for both spring and autumn phenology. Our results also outlined key challenges to address in further research to fully realise this potential and we hope that our publicly available code and data (link upon publication) will help these explorations towards robust phenology projection under climate change scenarios. 

\section*{Aknowledgements}
The work presented in this article was co-financed by the Federal Office for Meteorology and Climatology MeteoSwiss within the framework of GAW/GCOS-CH.

\section*{Author contribution}
Vivien Sainte Fare Garnot conceptualised the study, developed the deep learning methodology, ran the learning-based experiments, wrote the associated code, performed the formal analysis of the results, prepared the visualisations, and wrote the first version of the paper. Lynsay Spafford ran the process-models experiments , wrote the associated code, and helped in the formal analysis and the drafting of the first version of the paper. Jelle Lever contributed to the formal analysis and to the reviewing and editing of the manuscript.
Barbara Pietragalla and Chrisitian Sigg helped in the curation of the phenological data used in the present study. Barbara Pietragalla, Chrisitian Sigg, Yann Vitasse, Arthur Gessler, and Jan Wegner helped in the formal analysis and the editing of the manuscript. Arthur Gessler and Jan Wegner secured funding for the research.

\vspace{-2cm}

\begin{table*}[]
  \renewcommand{\arraystretch}{0.5}
    \caption{Leaf emergence process models included in our study, sorted in increasing complexity from top to bottom. }
    \centering
    \begin{tabular}{p{2.5cm} p{3.5cm} p{4cm} p{1cm} p{4cm} }
\toprule
\textbf{Model} & \textbf{Process(es) Included} & \textbf{Driver(s) Included} & \textbf{Free Parameters} & \textbf{Reference} \\
\midrule
Thermal Time (TT) & Ecodormancy Release & Forcing  & 3 & \cite{basler2016evaluating, reaumursebastopol, hufkens2018phenor, wang1960critique} \\
Photo-Thermal Time (PTT) & Ecodormancy Release & Forcing \& Photoperiod & 3 & 
\cite{basler2016evaluating, vcrepinvsek2006modelling, hufkens2018phenor, masle1989foliar} \\
Photo-Thermal Time with Sigmoidal Temperature Response (PTTs) & Ecodormancy Release & Forcing \& Photoperiod & 4 & \cite{basler2016evaluating, vcrepinvsek2006modelling, hanninen1990modelling, hufkens2018phenor, kramer1994selecting, masle1989foliar} \\
M1 & Ecodormancy Release & Forcing \& Photoperiod & 4 & \cite{basler2016evaluating, blumel2012shortcomings, hufkens2018phenor} \\
Alternating (AT) & Endodormancy \& Ecodormancy Release & Chilling \& Forcing & 5 & 
\cite{basler2016evaluating, cannell1983thermal, hufkens2018phenor, murray1989date} \\
Parallel (PA) & Endodormancy \& Ecodormancy Release & Chilling \& Forcing & 9 & 
\cite{basler2016evaluating, hanninen1990modelling, hufkens2018phenor, kramer1994selecting} \\
Sequential (SQ) & Endodormancy \& Ecodormancy Release & Chilling \& Forcing & 8 &
\cite{basler2016evaluating, hanninen1990modelling, hufkens2018phenor, kramer1994selecting, landsberg1974apple} \\
Dormphot (DP) & Dormancy Induction, Endodormancy \& Ecodormancy Release & Cooling, Chilling, Forcing, \& Photoperiod & 11 & \cite{basler2016evaluating, caffarra2011modelling, hufkens2018phenor} \\
\bottomrule

    \end{tabular}
    \label{tab:spring-pheno-process-models}
\end{table*}

\begin{table*}[]
\renewcommand{\arraystretch}{0.5}
    \caption{Leaf colouration process models included in our study. }
    \centering
    \begin{tabular}{p{2.5cm} p{3.5cm} p{4cm} p{1cm} p{4cm} }
    \toprule
\textbf{Model }& \textbf{Process(es) Included} & \textbf{Driver(s) Included }& \textbf{Free Parameters} & \textbf{Reference}\\
\midrule
White (WM) & Dormancy Induction & Cooling \& Photoperiod & 3 & \cite{liu2020modeling, white1997continental} \\
Delpierre 1 (DM1) & Dormancy Induction & Cooling \& Photoperiod & 5 &
\cite{liu2020modeling, zani2020increased, delpierre2009modelling} \\
Delpierre 2 (DM2) & Dormancy Induction & Cooling \& Photoperiod & 5 &
\cite{liu2020modeling, zani2020increased, delpierre2009modelling} \\
Delpierre 1-Zani 2020\\(DM1Za20) & Dormancy Induction & Cooling \& Photoperiod &
3 &
\cite{liu2020modeling, zani2020increased, delpierre2009modelling} \\
Jeong (JM) & Dormancy Induction & Cooling \& Photoperiod & 3 & 
\cite{liu2020modeling, jeong2014macroscale} \\
Dormphot Dormancy Induction (DPDI) & Dormancy Induction & Cooling \& Photoperiod & 4 & 
\cite{liu2020modeling, hufkens2018phenor, caffarra2011modelling} \\
Delpierre 1 with Preceding Spring Leaf Emergence (DM1s) & Preceding Ecodormancy Release \& Dormancy Induction & Forcing, Cooling, \& Photoperiod & 6 & 
\cite{liu2020modeling, delpierre2009modelling} \\
Dormphot Dormancy Induction with Preceding Spring Leaf Emergence (DPDIs) & Preceding Ecodormancy Release \& Dormancy Induction & Forcing, Cooling, \& Photoperiod & 5 &
\cite{liu2020modeling, caffarra2011modelling} \\
Photosynthesis-Influenced Autumn with Growing Season Index (PIAG) & Preceding Ecodormancy Release, Growing Season Productivity \& Dormancy Induction & Forcing, Cooling, Precipitation \& Photoperiod & 5 & 
\cite{meier2023process, zani2020increased} \\
\bottomrule
    \end{tabular}
    \label{tab:fall-pheno-process-models}
\end{table*}

\clearpage
\bibliography{00_climate-pheno}

\end{document}